\documentclass{article}
\usepackage[twocolumn]{emulateapj}
\usepackage{epsf}

\def\p{\phantom{1}}

\def\lae{\mathrel{<\kern-1.0em\lower0.9ex\hbox{$\sim$}}}
\def\gae{\mathrel{>\kern-1.0em\lower0.9ex\hbox{$\sim$}}}

\lefthead{{\sc C\^OT\'E ET AL.}}
\righthead{PALOMAR 13}
\submitted{Accepted for publication in the Astrophysical Journal}

\begin{document}

\title{Palomar 13: An Unusual Stellar System in the Galactic Halo\altaffilmark{1}}

\author{Patrick C\^ot\'e}
\affil{Department of Physics and Astronomy, Rutgers University, New Brunswick, NJ 08854; \\ pcote@physics.rutgers.edu}
\medskip

\author{S.G. Djorgovski}
\affil{California Institute of Technology, Mail Stop 105-24, Pasadena, CA 91125; \\ george@astro.caltech.edu}
\medskip

\author{G. Meylan\altaffilmark{2}}
\affil{European Southern Observatory, Karl-Schwarzschild-Strasse 2, 85748 Garching, Germany; and
Space Telescope Science Institute, 3700 San Martin Drive, Baltimore, MD 21218; \\ gmeylan@stsci.edu}
\medskip

\author{Sandra Castro}
\affil{California Institute of Technology, Mail Stop 105-24, Pasadena, CA 91125; \\ smc@astro.caltech.edu}
\medskip

\author{J.K. McCarthy}
\affil{California Institute of Technology, Mail Stop 105-24, Pasadena, CA 91125; and
PixelVision, 4952 Warner Avenue, Suite 300, Huntington Beach, CA 92649; \\ jkmccarthy@pacbell.net}
\medskip

\altaffiltext{1}{Based on data obtained at the W.M. Keck Observatory, which is operated as a scientific 
partnership among the California Institute of Technology, the University of California, and NASA, and was 
made possible by the generous financial support of the W. M. Keck Foundation.}

\altaffiltext{2}{Visiting Astronomer, Canada-France-Hawaii Telescope,
operated by the National Research  Council of Canada, the Centre National
de la Recherche Scientifique of France, and the University of Hawaii.}

\begin{abstract}
We report the first results of a program to study the internal kinematics of globular clusters
in the outer halo of the Milky Way. Using the Keck telescope and High Resolution Echelle Spectrometer, we have
measured precise radial velocities for 30 candidate red giants in the direction of
Palomar 13, an object traditionally cataloged as a compact, low-luminosity globular cluster.
We have combined these radial velocities with
published proper motion membership probabilities and new CCD photometry
from the Keck and Canada-France-Hawaii telescopes, to isolate a sample of 21 probable 
members. We find a systemic velocity of $\langle v_{r}\rangle_s = 24.1\pm0.5$ km s$^{-1}$
and a projected, intrinsic velocity dispersion of ${\sigma}_p = 2.2\pm0.4$ km s$^{-1}$.
Although small, this dispersion is nevertheless several times larger than that expected for 
a globular cluster of this luminosity and central concentration. Taken at face value, this
dispersion implies a mass-to-light ratio of $\Upsilon_V~ = 40^{+24}_{-17}$ based on
the best-fit King-Michie model.
The surface density profile of Palomar 13 also appears to be anomalous among
Galactic globular clusters --- depending upon the details of 
background subtraction and model-fitting, Palomar 13 either contains a substantial 
population of ``extra-tidal" stars, or it is far more spatially extended than previously 
suspected. The full surface density profile is equally well-fit by a King-Michie model having 
a high concentration and large tidal radius, or by a NFW model.
We examine --- and tentatively reject --- a number of possible explanations for the observed 
characteristics of Palomar 13 (e.g., velocity ``jitter" among the red giants, spectroscopic 
binary stars, non-standard mass functions, modified Newtonian dynamics), and conclude that the 
two most plausible scenarios are either catastrophic heating during a recent perigalacticon 
passage, or the presence of a massive dark halo. Thus, the available evidence 
suggests that Palomar 13 is either a globular cluster which is now in the process of dissolving
into the Galactic halo, or a faint, dark-matter-dominated stellar system.
\end{abstract}


\section{Introduction}

Much of our knowledge of the Galactic halo is based directly upon observations of 
its globular clusters, the most readily identifiable halo objects. Those clusters
which are located in the outer halo of the Milky Way are especially important probes 
of the formation and evolution of the Galaxy, as their ages and metallicities
provide direct constraints on the duration of halo formation process and on the 
time-scale for Galactic chemical enrichment, while the shape and extent of Galaxy's 
dark halo is constrained by their orbital properties.

These distant clusters share their location in the outer halo with a 
system of nine dwarf spheroidal (dSph) galaxies. These dwarfs have Galactocentric
distances in the range $24 \le R_G \le 250$ kpc (Mateo 1998) --- an interval which 
includes 16 globular clusters (Harris 1996). To some extent, the dSph galaxies resemble  
the globular clusters in that they always contain a population of old metal-poor stars;
there are, however, a number of important distinctions between the two classes of
objects. The central stellar densities of the dSph galaxies are much lower than those 
of typical globular clusters, and many of the dwarfs show evidence for intermediate-age 
populations (e.g., Mateo 1998). More importantly, measurements of the central velocity 
dispersions of the dSph galaxies suggest that they are embedded in extended dark halos 
(Aaronson 1983) which are presumably non-baryonic in nature (Feltzing, Gilmore \& Wyse 
1999), although alternative explanations have been proposed (e.g., Milgrom 1995; 
Klessen \& Kroupa 1998).

By contrast, color magnitude diagrams (CMDs) of distant halo globular clusters show 
no sign of distinct intermediate-age populations, although there is evidence that 
some of the clusters may be younger than their counterparts in the inner Galaxy 
(e.g., Stetson et~al. 1999). And while there is a preponderance of faint, 
low-density globular clusters in the outer halo (see, e.g., Figure 1 of van den Bergh 1994), 
their central stellar densities remain comfortably above those of even the most 
centrally concentrated dSph galaxies. 

Furthermore, dynamical studies of nearby globular clusters have established that, unlike the dSph 
galaxies, they appear to contain little or no dark matter apart from normal stellar 
remnants such as white dwarfs and neutron stars (e.g., Pryor \& Meylan 1993). However, 
a {\it direct} determination of the internal mass distribution in distant halo
clusters has proven impossible since the faintness of even their most evolved red giants 
--- coupled with their sparsely populated red giant branches (RGBs) and low
surface brightnesses --- make the requisite spectroscopic observations extremely challenging.
For example, 12 of the 16 globular clusters with $R_{G} \gae 24$ kpc have central surface 
brightnesses below ${\mu}_V \simeq 19.5$ mag arcsec$^{-2}$; for an assumed mass-to-light 
ratio\altaffilmark{3}\altaffiltext{3}{All mass-to-light ratios quoted
in this paper are in solar units.} of ${\Upsilon}_V = 2$, these clusters are expected to have 
central velocity
dispersions of ${\sigma}_{p,0} \lae 2$ km s$^{-1}$.  Thus, to study their internal 
dynamics, a velocity precision of $\epsilon(v_r) \lae 1$ km s$^{-1}$ for metal-poor stars in 
the range $17 \lae V \lae 20$ is required.

It is hardly surprising that these objects have been neglected in previous 
radial velocity surveys of Galactic globular clusters.  The lone exception is NGC 2419, for 
which Olzewski, Pryor \& Schommer (1993) measured velocities for 12 red giants. From this small sample,
they found ${\Upsilon}_V \simeq 0.7$, among the lowest measured for any globular cluster
and contrary to the notion that this cluster is embedded in a massive dark halo. 
Nevertheless, this object is --- with an absolute magnitude of $M_V = -9.6$ (Harris 1996) --- 
among the most luminous Galactic globular clusters and atypical of 
the clusters which populate the outer halo. Clearly, a dynamical study of a more representative 
sample of halo clusters is in order.

There are several reasons to suspect that a search for dark halos in outer halo
clusters might prove profitable. First, the characteristic mass of globular clusters 
is similar to the Jeans mass at recombination, and thus has a special cosmological 
significance; indeed, it has often been proposed that globular clusters may 
contain dark halos ($e.g.,$ Peebles 1984; West 1993). While there is no evidence
to support this conjecture, it has in practice proven surprisingly difficult 
to rule out (Heggie \& Hut 1996). For example, as clusters evolve in the Galactic tidal field,
they are expected to lose mass through tidal stripping ---
a process which confounds efforts to reconstruct the initial mass distribution 
in these clusters using present-day observations. Second, there are fairly compelling reasons
to believe that some Galactic dSph galaxies are dark-matter-dominated, and since globular clusters are
simply the next step down in luminosity, it is therefore reasonable to
expect that the same scenario which is usually invoked to explain the internal
dynamics of the dSph galaxies ($i.e.,$ gas ejection from shallow potential wells by 
supernova-driven winds, followed by adiabatic expansion; Dekel \& Silk 1986) might produce end-products 
that resemble globular clusters if the gas ejection mechanism was particularly efficient.
This possibility is especially topical in light of the emerging evidence from numerical
simulations that the halos of large galaxies have been assembled hierarchically
(e.g., Klypin et~al. 1999; Moore et~al. 1999 and 2001; 
C\^ot\'e et~al. 2000), although the predicted number of low-mass dark halos greatly exceeds
the observed number of dwarf galaxies in such simulations. Various means of resolving this important
contradiction have been proposed: i.e., by supressing the small-scale power in the family of cold dark 
matter models (see, e.g., Dav\'e et al. 2001), or by 
postulating the existence of a large population of ``dark galaxies"
(see, e.g., Hirashita, Takeuchi \& Tamura 1999; Bullock, Kravtsov \& Weinberg 2000).
There are also some reasons to believe that isolated, low-luminosity clusters may contain
significant amounts of {\it baryonic} dark matter. N-body simulations of evolving globular
clusters suggest that as clusters lose mass through dynamical evolution, the fraction of their 
mass contained in white dwarfs should increase steadily (Vesperini \& Heggie 1997). 
The outer halo contains a disproportionately high fraction of low-luminosity clusters,
so a search for dynamical evidence of extreme white dwarf populations in these clusters
might prove worthwhile.

In 1998, we began a program to study the internal dynamics of 
distant halo globular clusters using the High Resolution 
Echelle Spectrometer (HIRES) at the W.M.~Keck Observatory. This program, which was 
designed to yield the first direct measurements of the velocity dispersions and
mass-to-light ratios for these clusters, was also motivated by the need for 
improved measurements of the systemic velocities of several of these clusters
(i.e., the radial velocities of four of the 16 clusters having Galactocentric 
distances greater than 24 kpc are either unknown, or have uncertainties of
more than 20 km s$^{-1}$). It has been noted on numerous occasions that 
a number of the Galactic dwarf galaxies and globular clusters fall along great 
circles (e.g., Lynden-Bell 1976ab; Kunkel 1979; Majewski 1994).
Such structures would have obvious implications for the formation of the Galactic halo, but
unambiguous evidence of their existence requires orbital parameters
for the putative members.  Programs are now underway to measure the proper motions of 
the distant halo globular clusters and dwarf satellites (e.g., Dinescu et al. 2001), 
and accurate radial velocities are a prerequisite for the measurement of their orbits.

In this paper, we present the first results from this program. Our sample consists
of seven, predominantly low-concentration clusters having Galactocentric distances
in the range $25 \lae$ $R_G$ $\lae 112$ kpc. An analysis of the full sample
will be presented in a separate paper; here, we present our findings for a
single sparse object, Palomar 13.  With an absolute magnitude of
$M_V \simeq -3.8$ mag, Palomar 13 is exceeded in luminosity by all but a handful of 
the $\sim$ 150
globular clusters belonging to the Milky Way, and is one of the objects which has 
been identified on previous occasions as a possible member of a great stream 
(e.g., Lynden-Bell 1976a; Lynden-Bell \& Lynden-Bell 1995). Based on our
new radial velocities and photometry, we find both the structural properties and
the mass-to-light ratio of this object to be anomalous among the Galactic
globular cluster system.

\section{Observations and Reductions}

\subsection{Keck and CFHT Photometry}

On the night of 10 September, 1999, we used the Low Resolution Imaging Spectrometer 
(LRIS; Oke et~al. 1995) on the Keck II telescope to obtain a series of $BVI$ images of Palomar 13.
An observing log for these, and other, observations of Palomar 13 may be found 
in Table~\ref{tab1}. In imaging mode, LRIS has a spatial scale of $0\farcs215$ pixel$^{-1}$ 
and a field of view of 5\farcm8$\times7$\farcm3. For each filter, we obtained a pair of 
images centered on the cluster. Exposure times were 10 and 300s in $V$, 20 and 480s in $B$,
and 6 and 180s in $I$. Conditions during the night were photometric, and the FWHM of 
isolated stars within the $BVI$ frames were measured to be in the range 0\farcs7-0\farcs8.
Images were bias-subtracted and then flat-fielded in the 
IRAF\altaffilmark{4}\altaffiltext{4}{IRAF is distributed by the National Optical
Astronomy Observatories, which are operated by the Association of Universities for
Research in Astronomy, Inc., under contract to the National Science Foundation.}
environment using sky flats
obtained during twilight. Instrumental magnitudes for unresolved objects in
the field were derived using the DAOPHOT II software
package (Stetson 1993), and calibrated with observations of several Landolt
(1992) standard fields taken throughout the night. The calibration equations took
the form
\begin{equation}
\begin{array}{rcr}
V   & = & v +  a_1 - b_1X_V + c_1(b-v) \\
B-V & = & a_2 - b_2X_B + c_2(b-v) \\
V-I & = & a_3 - b_3X_I + c_3(v-i).\\
\end{array}{}
\label{eq1}
\end{equation}
Aperture corrections were determined separately for each filter, and a master object
list was created using the photometry files for the long and short exposures. 
The final photometric database contains 840 objects detected with a minimum point-source
signal-to-noise ratio of {\it S/N} = 5 in all three filters. A finding chart for Palomar 13 
constructed from our V images is shown in Figure~\ref{fig1}. All stars having measured 
radial velocities (see \S2.2) are identified on this image.

Images of Palomar 13 were also obtained on the night of 14 July, 1999
using the 3.6m Canada-France-Hawaii telescope (CFHT) and the CFH12K mosaic camera
(Cuillandre et~al. 2000). A series
of three 600s exposures in each of the $V$ and $R$ filters were taken with the cluster
centered on chip \#8 of the CCD mosaic. Each image was bias-subtracted and flat-fielded,
and then shifted and stacked to produce a final image for each of the two filters. Isolated
stars in the stacked images were found to have FWHMs of 0\farcs75 in each filter. 
Photometry was then performed using DAOPHOT II, and the photometry lists were matched.
Unfortunately, conditions were non-photometric and it was not possible
to calibrate the photometry directly. However, instrumental ({\it v}) magnitudes for stars on 
chip \#8 were calibrated via secondary standards in this field selected from 
our Keck/LRIS photometry. Thus, the photometric database from our CFHT data 
consists of calibrated $V$ magnitudes and instrumental $r$ magnitudes for 855 objects 
in this CCD field, which measures 7\farcm0$\times$14\farcm0. This single field 
is 2.3 times larger than that available from LRIS, although the photometric catalog is
limited to the range $18.5 \lae V \lae 23.5$ mag (as opposed to our LRIS 
point-source catalog, which spans the range $15.5 \lae V \lae 24.5$ mag).

By virtue of its remote location in the outer halo, Palomar 13 should, in principal, play
an important role in establishing the chronology of halo formation. Somewhat 
surprisingly, it has been the focus of only little attention in this regard. CMD studies 
are limited to photographic study of Ortolani, Rosino \& Sandage (1985), the 
CCD study of Borissova, Markov \& Spassova (1997) and the recent photographic/CCD study 
of Siegel et~al. (2001). In Figure~\ref{fig2}, we compare our LRIS magnitudes and 
colors for stars in common with each of these three studies. We find good agreement 
with the $BV$ photometry of Siegel et~al. (2001), but the comparison reveals some 
interesting discrepancies with the earlier studies. There is
evidence for a non-linearity in the $V$ magnitudes of stars fainter than $V \sim$ 21 
in the Ortolani et~al.  (1985) study, while the Borissova et~al. (1997) $V$ magnitudes
show a rather large scatter with a mean offset of ${\Delta}V$ = $0.22\pm0.06$ mag. Due to 
the small number of red giants in this cluster --- and the limited depth
of previous photometric studies --- the stars in common between 
these studies and ours span a modest range in color. For both the
Ortolani et~al. (1985) and Borissova et~al. (1997) datasets, we find mean 
color offsets of ${\Delta}(B-V)$ $\sim$ 0.1 mag. 

\subsection{HIRES Spectroscopy}

In a series of observing runs during 1998 and 1999, we used HIRES
(Vogt et~al. 1994) to acquire spectra for candidate 
red giants in the direction of seven distant halo globular clusters: Palomar 3, Palomar 4,
Palomar 5, Palomar 13, Palomar 14, NGC 7492 and NGC 2419. Table~\ref{tab1} presents
an observing log for our HIRES observations of Palomar 13, which were obtained during 
the course of four dark runs on the Keck I telescope. 

Candidate RGB stars in Palomar 13 were selected from previously published CMDs and 
finding charts (e.g., Ortolani et~al.  1985 and Borissova et~al. 1997). A few additional
targets were identified on the basis of instrumental CMDs constructed from images 
obtained with COSMIC (Kells et~al. 1998) on the Palomar 5m telescope.
During three of the four HIRES runs, we limited the entrance aperture to
0\farcs86$\times$7\farcs0 with the C1 decker and binned the 2048$\times$2048
detector 1$\times$2 (i.e., in the spatial direction) to reduce the read noise.
For the fourth run, we used the C5 decker (i.e., an entrance
aperture of dimension 1\farcs15$\times$7\farcs0) and binned the detector 2$\times$2.
The corresponding spectral resolutions for these two instrumental configurations are
${\lambda} / {\Delta}{\lambda} = 45,000$ and 34,000 respectively. In all cases,
we used a single readout amplifier, a gain setting of 2.4 $e^{-1}$ ADU$^{-1}$, the
red collimator and a cross-disperser in first order. The
angles of the grating and cross disperser were adjusted to give complete spectral
coverage over the range 5055 $\lae \lambda \lae$ 5355 \AA . Thorium-Argon 
comparison lamp spectra were acquired frequently during each night --- usually before 
and after each program star observation. In a few cases, lamp spectra were separated
by 30-60 minutes for short exposures of the brightest program stars. High-S/N spectra
for 8-12 IAU radial velocity standard stars were obtained during each observing run.

All spectra (i.e., program objects and standard stars) were reduced in an identical
manner following the general procedures described in earlier papers (e.g., C\^ot\'e
1999; C\^ot\'e et~al. 1999). The radial velocity of each program object was 
measured by cross-correlating its spectrum against that of a master template
created during each run from the observations of IAU standard stars. In
order to minimize possible systematic effects, a master template for
each observing run was derived from a similar, and in some cases identical, sample 
of IAU standard stars. From each cross-correlation function, we measured both 
$v_r$, the heliocentric radial velocity, and $R_{TD}$, the Tonry \& Davis (1979)
estimator of the strength of the cross-correlation peak. During the course of
this program --- which spanned seven observing runs totaling 13 nights ---
we obtained 53 distinct radial velocity measurements for 23 different program
objects. All of these objects are faint, metal-poor members of our program 
clusters. Using the procedures described in Vogt et~al. (1995) and C\^ot\'e et~al. (1999), 
we then derived empirical estimates for our radial velocity uncertainties. 
Specifically, we assume that the uncertainty of any radial
velocity measurement, ${\epsilon}(v_r)$, can be expressed
\begin{equation}
{\epsilon}(v_r) = {\alpha}/(1 + R_{TD})
\label{eq2}
\end{equation}
where ${\alpha}$ is a constant to be determined. For our sample of repeat 
velocity measurements, which has 30 degrees of freedom, we calculate 
${\alpha} = 9.0^{+2.4}_{-1.6}$ for $\chi^2 = 1$, where the
quoted uncertainties refer to 90\% confidence limits. 

Table~\ref{tab2} summarizes the results of our photometric and spectroscopic observations
of Palomar 13. From left to right, this table records the name of each program 
star,\altaffilmark{5}\altaffiltext{5}{Identifications are from Ortolani et~al. (1985)
for those stars beginning with either a ``ORS" or ``V" prefix; four additional stars 
(\#910, 911, 915 and 931) which were not included in the Ortolani et~al. (1985) study 
were selected from images taken with COSMIC on the Palomar 5m telescope.}
distance from the cluster center, $V$ magnitude, $(B-V)$
color, HIRES exposure time, heliocentric Julian date, Tonry \& Davis $R_{TD}$ value,
heliocentric radial velocity, and the weighted mean velocity. The final column 
gives the cluster membership probability, P($\mu$), taken
from the proper motion survey of Siegel et~al. (2001). These membership probabilities 
are discussed in detail below.

\section{Results}
\subsection{Identification of Members and Non-members}

A crucial first step in studying the internal dynamics of Palomar 13 is the
isolation of a sample of {\it bona fide} members. At a Galactic latitude of
$b = -42\fdg7$, Palomar 13 is located well below the Galactic plane, but
its very sparse giant branch\altaffilmark{6}\altaffiltext{6}{Ortolani 
et~al. (1985) reported that the cluster luminosity function is similar to that of M3, 
but reduced by a factor of $\sim$ 1/60.} means that contamination from foreground disk 
and halo stars may be non-negligible. As a further complication, there is only limited 
radial velocity separation with the disk field star population due to
Palomar 13's low systemic velocity (see \S 3.5). These difficulties 
are ameliorated by the high precision of our radial velocity measurements, our new 
CMD, and the availability of proper motions from Siegel et~al. (2001).

In Figure~\ref{fig3}, we plot the mean radial velocities of our program stars, 
$\langle v_r\rangle $, against membership probability, P(${\mu}$), from Siegel et~al. 
(2001).\altaffilmark{7}\altaffiltext{7}{Two stars have been omitted: V2 (a known RR Lyrae; 
Ciatti et~al. 1965) and \#915. Siegel et~al. (2001) were unable to measure proper motions 
for either of these two objects, but both stars are almost certainly members based on
their radial velocities, their location in CMDs, and their proximity to the 
center of Palomar 13 (see columns 2 of Table~\ref{tab2} and Figure~\ref{fig1}). 
Nevertheless, V2 has been excluded from the calculation of the observed
velocity dispersion since atmospheric pulsations may inflate the measured dispersion 
(e.g., the peak-to-peak radial velocity variations of RR Lyrae stars in M92 
are ${\Delta}v_r = 50-60$ km s$^{-1}$; Cohen 1992).} The sharp spike at 
$\langle v_r\rangle$ $\sim$ 25 km s$^{-1}$ identifies Palomar 13; the vertical dashed
line indicates our best estimate for its systemic velocity (see \S 3.5). In general, there 
is good agreement between the membership classifications based on the radial velocities 
and those based on the proper motions. The three exceptions are ORS-23, ORS-32 and ORS-118. In the first
case, the measured radial velocity of $v_r = 18.57\pm0.82$ km s$^{-1}$ for ORS-23 provides
no compelling evidence either for, or against, membership. However, 
Siegel et~al. (2001) find P(${\mu}$) = 0\% for this star, a conclusion which is
supported by the location of this star in the CMD: i.e., the star
has a very red color of $(B-V)$ = 1.50 mag and is located $\sim$ 0.8 mag blueward
of Palomar 13's RGB. We conclude that ORS-23 is a likely field star.
For ORS-32, Siegel et~al. (2001) report an intermediate probability of P(${\mu}$) = 35\%. Our
mean radial velocity, $v_r = 19.68\pm0.26$ km s$^{-1}$, is based on four measurements at
three different epochs. The residual with respect to the systemic velocity of Palomar 13 is 
4.5 km s$^{-1}$, and the star is located $R \sim 1^{\prime}$ from the core of Palomar 13.
We believe that these facts, coupled with the position of this star in the CMD 
(particularly in the $V,~V-I$ plane), marginally favor the interpretation that it
is a member of Palomar 13. Nevertheless, a spectroscopic measurement of the metallicity 
of this star would be desirable.

The situation for ORS-118 is somewhat ambiguous. Although Siegel et~al. (2001) report
P(${\mu}$) = 0\% for this star, its location in the CMD strongly suggests that it is 
a true member, since it lies precisely on the RGB of the best-fit isochrone (see \S 3.3).
The measured radial velocity of this star, $v_r = 24.92\pm0.21$ km s$^{-1}$, is
indistinguishable from the systemic velocity of Palomar 13, and is based on three
independent measurements which agree internally to better than 0.75 km s$^{-1}$.
We further note that an
abundance analysis for ORS-118 (described below) yields [Fe/H] = $-1.98\pm0.31$ dex. 
This metallicity points to a halo origin, and is consistent with the metallicity
of Palomar 13 deduced from isochrone fitting (see \S 3.3). 

To investigate the possibility that ORS-118 is simply an interloping foreground star 
which happens to have the same radial velocity as the Palomar 13, we approximate 
the line-of-sight velocity distributions expected for disk and halo stars in this
direction with the relation
\begin{equation}
{\rm P(}v_r{\rm )} \propto \exp{[-(\langle v_r\rangle  - \langle v_{r,0}\rangle)^2/2{\sigma}^2}]
\label{eq3}
\end{equation}
where
\begin{equation}
\begin{array}{rcll}
\langle{v_{r,0}}\rangle & = & -19.5\cos{\phi_d}  & \\
\end{array}
\label{eq4}
\end{equation}
is the mean radial velocity of disk stars along this line of sight as a result of the solar motion
towards $l = 56^{\circ}$ and $b = 23^{\circ}$ (Mihalas \& Binney 1981). The angle between this point
on the celestial sphere and Palomar 13 is denoted by $\phi_d$.
Similarly, the mean radial velocity of halo stars along this 
line of sight is taken to be
\begin{equation}
\begin{array}{rcll}
\langle{v_{r,0}}\rangle & = & -220\cos{\phi_h} &  \\
\end{array}
\label{eq5}
\end{equation}
where $\phi_h$ is the angle subtended by Palomar 13 and ($l,~b$) = (90$^{\circ}$, $0^{\circ}$).
We further approximate the local velocity dispersion of disk stars as 
\begin{equation}
{\sigma}^2 \simeq {1 \over 3}({\sigma}_U^2 + {\sigma}_Y^2 + {\sigma}_Z^2 )
\label{eq6}
\end{equation}
which, from the last row of Table 1 of Dehnen \& Binney (1998), yields ${\sigma} \simeq 29$ km s$^{-1}$.
For the halo, we assume an isotropic velocity ellipsoid (i.e., 
${\sigma}_U = {\sigma}_V = {\sigma}_Z$) and adopt ${\sigma} \simeq 124$~km~s$^{-1}$
from C\^ot\'e (1999).

The resulting probability distributions for disk and halo stars are illustrated
by the dotted lines in Figure~\ref{fig3}. Although the absolute normalization of the two
curves is arbitrary, their {\it relative} normalization has been adjusted to match the
expected number of disk and halo stars with $16.5 \le V \le 20$ along this line of
sight (determined from the IAS Galaxy model; Bahcall \& Soneira 1980). If ORS-118 is
indeed a halo field star, then it lies 1.5$\sigma$ from the mean of the field distribution;
the probability of such an occurrence is 13\%.
Moreover, in the range $16.5 \le V \le 17.5$,
the IAS Galaxy model predicts only a single halo field star in our field, and 
just 8\% of halo stars in this magnitude range are expected to have 
$(B-V) \ge 0.93$ (i.e., the color of ORS-118). We
conclude that ORS-118 is likely to be a {\it bona fide} member of 
Palomar 13. The origin of the large proper
motion residual measured by Siegel et~al. (2001) is unclear, particularly since an 
inspection of our CCD images revealed no close companions.

\subsection{Metallicity}

Exposure times for our HIRES observations were chosen to yield the minimum signal-to-noise needed 
to derive radial velocities with a precision ${\epsilon}(v_r) \lae 1$ km s$^{-1}$. For ORS-118, 
the brightest star in our sample which has a radial velocity consistent with membership in 
Palomar 13, we combined the two spectra obtained during the July 1999 observing run. The
signal-to-noise of ratio of this co-added spectrum --- $S/N \simeq 25$ per resolution element ---
is just adequate for abundance analysis.

Our analysis used 28 Fe I lines with oscillator strength values adopted from McWilliam et~al. (1995),  
Kurucz model atmospheres and solar abundances from Anders \& Grevesse (1989). Computations 
were made with the most recent version of the LTE line analysis code MOOG (Sneden 1973). 
The effective temperature, $T_{\rm e}$, was first determined using an initial estimate
from the photometry and then checked through excitation equilibrium of the Fe I lines whose
equivalent widths were measured using standard IRAF routines. The surface gravity, $\log{g}$, 
was derived from the ionization equilibrium of Fe I and Fe II lines (using four Fe II lines 
in the available spectral region). Once the effective temperature and surface gravity were 
determined, abundances for the Fe I lines were calculated by iteratively varying the 
microturbulent velocity until the best fit in the diagram of Fe I abundance versus 
equivalent width was achieved. Our best-fit parameters for ORS-118 are $T_{\rm e}$ = 4700 K, 
$\log{g}$ = 1.7, and [Fe/H] = $-1.98\pm0.31$ dex, where the rather large uncertainty 
on the measured metallicity is a reflection of the modest signal-to-noise of our co-added spectrum.

Previous estimates for the metallicity of Palomar 13 include [Fe/H] = $-1.9\pm0.4$
(Canterna \& Schommer 1978), $-1.67\pm0.15$ (Zinn \& Diaz 1982) and $-1.9\pm0.1$ dex
(Friel et~al. 1982). The measurement of Canterna \& Schommer (1978) comes from
Washington photometry of individual stars, while the latter two measurements are
based on low-resolution spectroscopy. Our determination of the cluster metallicity
is slightly lower than, but still consistent with, these previous estimates. In
the following section, we show that this metallicity is also consistent with that
found from isochrone fitting of the CMD.

\subsection{Color Magnitude Diagram}

Figure~\ref{fig4} shows $BV$ and $VI$ CMDs for Palomar 13 constructed from our LRIS 
photometry. As noted by previous investigators, the RGB is very sparsely populated.
This figure (and the left panel of Figure~\ref{fig6}) also confirms the finding of 
Siegel et~al. (2001) that Palomar 13 contains a population of blue stragglers;
likwise, there may be evidence for a ``second sequence" which runs parallel to the 
upper main sequence. Both this latter sequence (which is traditionally 
interpreted as the signature of unresolved binary stars; e.g., Romani \& Weinberg 1991),
and the presence of blue stragglers (a fraction of which are likely to be W~UMa 
systems; Mateo et~al. 1990) provide some evidence
for a population of binary stars. If this is indeed the case, then it would have 
important implications for modeling the observed velocity dispersion profile. In
\S 5.1, we shall return to the question of binary stars and their possible effect on the
observed velocity dispersion.

From the DIRBE maps of Schlegel, Finkbeiner \& Davis (1998), we find the
reddening in the direction of Palomar 13 to be $E(B-V)$ = 0.11 mag.
The solid curve in the left panel of Figure~\ref{fig4} shows a 14 Gyr isochrone from 
Bergbusch \& VandenBerg (1992) with [Fe/H] = $-1.78$ dex, shifted by $(m-M)_V = 17.27$ mag. 
We stress that the sparsely populated red giant and horizontal branches preclude a precise 
age determination, and simply note that a 14 Gyr isochrone provides a reasonable match to 
the observed fiducial sequences.
Assuming $A_V = 3.1E(B-V)$, we find a true distance modulus of 
$(m-M)_0 = 16.93\pm0.10$ mag, which corresponds to a distance of $D = 24.3^{+1.2}_{-1.1}$ kpc.
This value is in good agreement with the distance of $D = 24.8$ estimated by
Siegel et~al. (2001) from the average magnitude of a small sample of cluster
RR Lyrae and horizontal branch stars. 
Based on the combination of isochrone fitting to the cluster CMD and the
spectroscopic analysis of ORS-118, we adopt [Fe/H]=$-1.9\pm0.2$ dex as our best
estimate for the metallicity of Palomar 13.

Ortolani et~al. (1985) found an absolute magnitude of $M_V = -3.4$ mag for Palomar 13
by integrating the luminosity function. For a distance of $D = 24.3$, 
we find an integrated luminosity of $2.4\times10^{3}~L_{V,{\odot}}$ ($M_V = -3.6$) within 
the tidal radius suggested by Siegel et al. (2001), $r_t \simeq 188^{{\prime}{\prime}}$
(although, as discussed in \S 3.4, we see no evidence for a tidal cutoff 
at this, or any other, radius).
This estimate excludes stars with $(B-V)$ $<$ 1.3 mag and confirmed 
non-members. If the integration is extended over the entire LRIS field of view
(as would be appropriate for the larger tidal radius found in \S 4), the luminosity
increases to $3.1\times10^{3}~L_{V,{\odot}}$ ($M_V = -3.9$). Our corresponding best 
estimate for the absolute magnitude of Palomar 13 is $M_V = -3.8$ mag (see also \S 4-5).
In the most recent version of
the Harris (1996) catalog of Galactic globular clusters, only three objects --- E 3, 
Palomar 1 and AM 4 --- have lower luminosities.\altaffilmark{8}\altaffiltext{8}{According
to Harris (1996), the absolute magnitude of Terzan 1 is $M_V = -3.3$; i.e.,
fainter than Palomar 13. However, Idiart et~al. (2002) have recently reported an upward
revision to $M_V \simeq -5.4$ for this cluster.}
The integrated, de-reddened color is ${\langle B-V\rangle}_0 = 0.65\pm0.04$ mag, roughly 
consistent with that expected for a globular cluster of metallicity [Fe/H] $\simeq -1.9$ 
dex (Durrell et al. 1996). 

\subsection{Surface Density and Surface Brightness Profiles}

While there are surprisingly few published studies of the structural properties
of Palomar 13, previous results have consistently pointed to a compact, faint cluster of low 
concentration. For instance, Webbink (1985) quotes a core radius of $r_c = 23^{\prime\prime}$ 
and a concentration index of $c \equiv \log{(r_t/r_c)} = 0.9$, while Trager et~al. (1995) 
report $r_c = 29^{\prime\prime}$ and $c = 0.66$. However, both of these studies relied primarily 
on the star counts of Kinman \& Rosino (1962) which were obtained by viewing the projection of 
a photographic plate through a mask fixed on the screen of a Sartorious astrophotometer.
Recently, Siegel et~al. (2001) adopted $c = 0.7$ and measured 
$r_c = 39^{\prime\prime}$ for Palomar 13 using a sample of 
119 members selected on the basis of their proper motions, but noted that this model
provides a poor description of the measured profile at large radii: i.e., 
14 of their proper motion members (12\% of the sample)
lay beyond the canonical tidal radius. This fact, coupled with Palomar 13's low luminosity
and eccentric orbit, led Siegel et~al. (2001) to suggest that this object is in the final
stages of destruction.

We have used our Keck and CFHT imaging to measure improved structural parameters for
Palomar 13. The point-source catalog from our Keck images consists of 841 objects
distributed over an area of 42.8 arcmin$^2$. Since the Palomar 13 field 
contains a significant number of obvious foreground/background objects --- see 
Figure~\ref{fig4} --- the Keck point-source catalog was trimmed to exclude objects 
with $0.0 < (B-V) < 1.3$ mag. Similarly, sources which did not fall in the range 
$0.15 < (V-r) < 0.7$ were culled from the CFHT catalog. 
The point-source catalogs were further restricted to objects brighter than $V = 24$ (Keck)
and 23.5 (CFHT), in order to ensure photometric completeness at all radii. This
leaves a total of 629 and 551 objects in the Keck and CFHT databases, respectively.

Figure~\ref{fig5} shows the surface density profiles determined from these catalogs. 
Observations from Keck are shown as open squares, while open circles denote CFHT data 
points.  The latter have been scaled upward by a factor (629/551) $\simeq$ 1.14 to account 
for the difference in sample size. There is good agreement between these two independent 
profiles. Given the larger radial extent of the CFHT data, they provide stronger constraints 
on the asymptotic 
behavior of the surface density profile --- a particularly important issue in the present case since the
Keck surface density profile continues to decline out to the edge of the LRIS field of view. 
The arrow in Figure~\ref{fig5} shows our best estimate for the background surface density, 
${\Sigma}_b = 0.68$ stars arcmin$^{-2}$, based on the scaled CFHT observations. For
comparison, the surface density of Galactic stars in the appropriate color and magnitude
intervals in the direction of Palomar 13 predicted by Institute for Advanced Study Galaxy 
model (Bahcall \& Soneira 1980; Bahcall 1986) fall in the range $0.64 \lae {\Sigma}_b \lae 0.95$
stars arcmin$^{-2}$, consistent with our empirical estimate.
The filled symbols in this figure show the background-subtracted surface density profile.

For comparison, the open triangles in Figure~\ref{fig5} shown the surface density profile of 
Siegel et al. (2001), taken directly from a digitized image of their Figure 7. The Siegel
et al. data points have been scaled upwards by the ratio of the numbers of stars in the
two studies: (629/119) $\simeq$ 5.29. In the inner regions of Palomar 13, the agreement
between the two profiles is excellent. Beyond a distance of $R = 2-3^{\prime}$, though,
the profiles differ, with our surface densities being higher by a factor of $\sim$ five
(although, to within the rather large uncertainties, the {\it slopes} of the two profiles 
are in agreement).  The origin of this difference is unclear. The profile of Siegel et al. 
(2001) has the obvious advantage of being proper motion selected: i.e., objects having 
membership probabilities of P(${\mu}$)~$<$~50\% have been rejected on a star-by-star basis. 
In principal, the use of proper motions to cull interlopers should be superior to the 
{\it statistical} approach used here to estimate the background level. On the other
hand, Siegel et al. (2001) note that most of their plates extend to only $V \sim 21$ (i.e.,
the approximate location of the main-sequence turnoff) whereas each of our two independent sets of
CCD photometry reach several magnitudes below this level, and with small photometric errors. 
If Palomar 13 is experiencing significant mass loss through evaporation and tidal stripping 
(see \S 5.2), then it is the lowest mass stars which are expected to have largest evaporation
rates (i.e., they should preferentially inhibit the cluster envelope as a result of mass 
segregation). In such a case, our deeper star counts compared to Siegel et al. (2001) might
lead to a more pronounced surface density excess at large radii. 
In any event, we caution that the five outermost bins in the Siegel et al. (2001) profile 
(spanning the range $2^{\prime} \lae R \lae 10^{\prime}$) contain just 15 stars, so 
it is unclear of the difference between the two studies is truly significant. 

With these caveats in mind, we point out a striking feature of Figure~\ref{fig5} --- the near power-law 
behavior of the surface density profile. Indeed, a single power-law, 
${\Sigma}$($R$) $\propto$ $R^{-{\gamma}}$,
with $\gamma = 1.8\pm0.2$ provides an adequate description of the surface density profile 
for $R \gae$ 0\farcm3. 
The dotted curve in Figure~\ref{fig5} shows the King-Michie model proposed by Siegel 
et al. (2001). This model, which has a somewhat arbitrary concentration parameter of $c = 0.7$ 
and a core radius of $r_c = 39^{{\prime}{\prime}}$, 
provides adequate description of the observed surface density profile over the range
0\farcm3 $\lae R \lae$ 1$^{\prime}$, but underestimates the central surface density.
More importantly, if this model accurately reflects the true density profile of Palomar 13, 
there is an even more dramatic ``excess" of stars above the model predictions beyond
$R~\gae$~1$^{\prime}$. While it is certainly true that the exact 
shape of the surface density profile depends on the adopted background level
(particularly at large radii) this excess is unlikely to be an
artifact of the background subtraction. For instance, achieving an acceptable
fit to a King model with $c \lae 1$ for $R \lae$ 2$^{\prime}$ would require
the background level to be increased by roughly an order of magnitude.

The dashed curve in Figure~\ref{fig5} shows the best-fit King-Michie model, which will
be discussed in detail in \S 4. As shown there, 
the King-Michie which best fits the observed surface
density and surface brightness profiles of Palomar 13 has a tidal radius of 
$r_t = 26\pm6^{\prime}$ (182$\pm$41 pc). Since this is more than twice the radial 
extent of our CFHT imaging, the adopted background level should be viewed
with some caution. However, we note that estimating the background surface density 
from star counts made within $r_t$ should result in an {\it overestimate} of the 
true background level and, hence, to an {\it underestimate} of $r_t$.
{\it We are therefore left with two possible interpretations of the surface
density profile of Palomar 13: either this object contains a significant population 
of ``extra-tidal" stars which have been caught in the act of evaporation (as suggested 
by Siegel et al. 2001), or its concentration and spatial extent have been greatly
underestimated in previous studies.} 

We now turn our attention to the properties of the stars associated with this 
surface density excess. This feature is evident in {\it both} the CFHT and LRIS surface 
density profiles, but is there any evidence that this population is truly comprised of Palomar 13 members?
Figure~\ref{fig6} shows instrumental CMDs for the Palomar 13 field derived from our CFHT 
photometry, along with the color and magnitude limits used to cull probable non-members
from the sample (i.e., the dotted regions). The panel on the left shows the CMD for all 
stars within the Siegel et al. (2001) tidal radius. The middle panel shows the CMD for all objects
beyond this radius, while the right panel shows those stars located at more than twice
this distance. The dashed curves in the final two panels show the Palomar 13 fiducial
sequence. A significant number of probable members with $R > r_t$ are evident, providing
additional evidence that the stars belonging to the excess in the surface density
profile are indeed associated with Palomar 13.

Figure~\ref{fig7} shows the surface {\it brightness} profile for Palomar 13 measured from
our Keck and CFHT images. Profiles having been calculated independently using both the Keck 
and CFHT photometry. Surface photometry is shown as the pentagons (Keck)
and triangles (CFHT), while the results from star counts are indicated by the squares 
(Keck) and circles (CFHT). A direct measurement of the background surface brightness
for the Keck and CFHT images yields $\mu_V = 21.3$ and 21.5 mag arcsec$^{-2}$,
respectively --- consistent with expectations for dark sky conditions at Mauna Kea
during 1999 (Krisciunas 1997). In deriving a surface brightness profile from the star 
counts discussed above, we have simply multiplied the surface densities 
shown in Figure~\ref{fig5} by a constant factor, $\eta = L_V/n_{\rm tot}$, where
$L_V$ is the integrated luminosity of Palomar 13 and $n_{\rm tot}$ are the number of 
stars used to derive each of the Keck and CFHT surface density profiles. This
approach suppresses large fluctuations in the surface brightness profile caused
by the small number of red giants in this low luminosity object.

How does our surface brightness profile compare with previous findings? Unfortunately,
Ortolani et al.  (1985) and Siegel et al. (2001) each presented surface density profiles for 
Palomar 13, but neither reported surface brightness measurements. Webbink (1985) quotes a 
central surface brightness of $\mu_V$ = 22.41 mag arcsec$^{-2}$ based on the star counts 
of Ortolani et al. (1985) and the aperture photometry of Racine (1975). Trager et al. (1995) 
found a similar core radius and concentration (i.e., $r_c = 29^{{\prime}{\prime}}$
and $c = 0.66$) but a much different central surface brightness: 
$\mu_V$ = 24.31 mag~arcsec$^{-2}$. The overall shape of the Trager et al. (1995) surface brightness 
profile, shown by the open stars in Figure~\ref{fig7}, is in reasonable agreement with 
the profile found here, but differs systematically by a factor of $\sim$ 3.4 in surface 
brightness (in the sense that their measurements are lower).
Their profile consists primarily of star counts Kinman \& Rosino
(1962), normalized to the aperture photometry of Racine (1975). We believe that
the normalization given here is to be preferred since two independent measurements
(from Keck and CFHT) are in close agreement. Moreover, an integration of the best-fit
King-Michie model of Trager et al. (1995) yields an absolute magnitude for Palomar 13
of $M_V = -1.25$ mag. This is $\sim$ ten times fainter than the value of 
$M_V = -3.6$ reported by Ortolani et al. (1985) and the value of $M_V = -3.8$ found here,
both of which were obtained by direct integration of the luminosity function.

The various curves shown in Figure~\ref{fig7} represent different models for the
light distribution in Palomar 13. We will discuss these models in detail below (in \S 4-5),
but before doing so, we shall use the radial velocities presented in Table 2
to investigate the internal kinematics of Palomar~13.

\subsection{Internal Kinematics}

Using our sample of 21 members (V2 excluded), we calculate a systemic 
velocity of $\langle v_r\rangle_s = 24.1\pm0.5$ km s$^{-1}$ using the maximum likelihood
estimator of Pryor \& Meylan (1993). The corresponding estimate for the
the projected {\it intrinsic} velocity dispersion using this same sample is 
${\sigma}_p = 2.2\pm0.4$ km s$^{-1}$. Omitting ORS-118 from the
sample (see \S 3.1) has essentially no effect on the derived parameters:
$\langle v_r\rangle = 24.0\pm0.5$ km s$^{-1}$ and ${\sigma}_p = 2.3\pm0.4$ km s$^{-1}$.
Likewise, if ORS-32 is omitted from the sample, the best-fit parameters do not
change significantly: $\langle v_r\rangle = 24.3\pm0.5$ km s$^{-1}$ and 
${\sigma}_p = 2.0\pm0.4$ km s$^{-1}$.

In the upper panel of Figure~\ref{fig8}, we show the radial velocities for these 
stars plotted against distance from the center of Palomar 13. The horizontal
dashed line shows the systemic velocity, $\langle v_r\rangle_s = 24.1$ km s$^{-1}$,
while the vertical arrows at $R = 1^{\prime}$ denote the radius where the possible 
excess in the surface density profile begins. It is perhaps noteworthy that
the stars with the largest velocity residuals are located at, or beyond, this radius. 
We note that there is a tendency for stars with the most discrepant velocities to also 
have the smallest uncertainties --- this is a consequence of the fact 
that these stars were observed on multiple observing runs to confirm the first-epoch 
velocities and to search for any evidence of binarity. As with the full sample, none
of these stars shows compelling evidence for radial velocity variations.

In the lower panel of Figure~\ref{fig8}, we plot the velocity dispersion profile
for Palomar 13. The open circle denotes the dispersion found using our complete sample 
of 21 members, plotted at their mean radius: $\langle R\rangle$ = 0\farcm84 (5.9 pc).
The filled circles show the results of dividing the sample into two radial bins 
containing 11 and 10 stars, respectively. In this case, the dispersions are
1.0$\pm$0.4 km s$^{-1}$ at $\langle R\rangle$ = 0\farcm36 (2.6 pc) and
2.6$\pm$0.6 km s$^{-1}$ at $\langle R\rangle$ = 1\farcm36 (9.6 pc). 
Although there may be some evidence for a rising velocity dispersion profile, we caution
that this result is based on a small sample, and that the uncertainties in the
measured dispersions are appreciable. 

How does the measured velocity dispersion for Palomar 13 compare with expectations?
Globular clusters are well described by King-Michie models, in which the central velocity 
dispersion (in km~s$^{-1}$) may be approximated by
\begin{equation}
{\sigma}_{p,0}^2 \simeq 0.003{\Upsilon}_V{r_c}{\mu_{V,0}} \\
\label{eq7}
\end{equation}
where ${\Upsilon}_V$ is the $V$-band mass-to-light ratio, $r_c$ is the core radius in pc, and 
${\mu_{V,0}}$ is the central surface brightness is units of $L_{V,{\odot}}$~pc$^{-2}$
(e.g., Richstone \& Tremaine 1986; Mateo et~al. 1991).
Based on the King-Michie model fits to the observed surface brightness profile presented in 
\S 3.4 and Table~\ref{tab3}, the {\it expected} velocity dispersion of Palomar 13 is
$\sigma_{p,0} \simeq 0.6$~km~s$^{-1}$ for 
an assumed mass-to-light ratio of ${\Upsilon}_V = 2$.  Thus, our measured velocity 
dispersion exceeds the expected value by roughly a factor of four.  The implications 
of this finding are presented below, along with a more detailed analysis of the 
internal kinematics Palomar~13.

\section{Mass-to-Light Ratio}

As discussed in \S 3.4, our new surface density profile for Palomar 13 suggests that either
a significant fraction of its stars are located beyond the ostensible tidal radius, or
the object is more spatially extended than previously believed. We now examine its internal
kinematics within the context of these two scenarios.

The short-dashed curve in Figure~\ref{fig7} shows the King-Michie which best fits the 
observed surface brightness profile for Palomar 13. The parameters for this model are 
$r_c = 14\pm2^{\prime\prime}$, $c = 2.0\pm0.15$ and 
$\mu_{V,0} = 22.54^{+0.30}_{-0.17}$~mag~arcsec$^{-2}$. (These
and other model parameters are summarized in Table~\ref{tab3}, along with various observed and
derived properties of Palomar 13.) 
According to this model, Palomar 13 has both a higher concentration {\it and} a greater
spatial extent ($r_t \sim 180$ pc) than suggested by previous studies. Indeed, this
concentration index is considerably larger than that expected for a globular cluster of
this luminosity or surface brightness (see, e.g., Djorgovski 1991). Integration of 
this profile yields a total luminosity of
$L_V \simeq (2.8\pm0.4)\times10^3$ $L_{V,{\odot}}$. The corresponding velocity dispersion profile
for an assumed mass-to-light ratio of ${\Upsilon}_V = 2$ is shown by the lower dashed curve
in the bottom panel of Figure~\ref{fig8} (i.e., curve 1-A). In agreement with the naive
analysis presented above, 
we find the measured velocity dispersion to exceed that expected for a normal globular
cluster of this concentration and luminosity. A maximum likelihood solution for the best-fit scale
velocity of this same King-Michie model gives the upper dashed curve shown in this figure
(i.e., curve 1-B). This model has a central velocity dispersion of $\sigma_{p,0} \simeq 2.5$
km~s$^{-1}$, and a total mass of $M = (1.10^{+0.44}_{-0.36})\times10^5$ $M_{\odot}$. The mass-to-light
ratio is then ${\Upsilon}_V = 40^{+24}_{-17}$ (1$\sigma$ uncertainties).

Alternatively, we may use the King-Michie model suggested by Siegel et al. (2001) to estimate
the mass and mass-to-light ratio of Palomar 13. This model, which is shown by the dotted curve in
Figure~\ref{fig7}, yields a total luminosity of $L_V \simeq 1.2\times10^3$ $L_{V,{\odot}}$.
A maximum likelihood scaling of the model velocity dispersion profile to the measured
velocities produces the dotted curve in the lower panel of Figure~\ref{fig8} (i.e., curve 2).
For this model, which has a central velocity dispersion of $\sigma_{p,0} \simeq 2.8$ km~s$^{-1}$,
the mass within the tidal radius ($r_t = 188^{\prime\prime}$ or 23 pc) is
$M = (5.5^{+1.9}_{-1.6})\times10^4$ $M_{\odot}$. The corresponding mass-to-light ratio is 
${\Upsilon}_V = 48^{+17}_{-14}$ where the quoted 1$\sigma$ uncertainties refer only to the errors
on the derived mass. {\it Thus, irrespective of the adopted surface brightness profile, it is not
possible to reproduce the observed velocity dispersion with 
a mass-to-light ratio that is typical of globular clusters 
(e.g., ${\langle\Upsilon}_V\rangle = 2.3\pm1.1$; Pryor \& Meylan 1993).}

\section{Implications}

What are the implications of this finding? 
We begin by examining --- and tentatively rejecting --- the possibility that the observed 
velocity dispersion has been inflated by a ``velocity jitter" in the atmospheres of the program 
stars, or by a population of spectroscopic binary stars. We then examine other
scenarios which may explain the observed properties of Palomar 13 including
tidal disruption, a mass function which is heavily skewed to low-mass stars or heavy 
stellar remnants, modified Newtonian dynamics, and/or the presence of a massive dark halo.
It is worth bearing in mind that these various possibilities may not be
mutually exclusive (i.e., a very high binary fraction combined with tidal heating
and mass loss).

\subsection{Velocity Jitter and/or Binary Stars?}

The value of ${\sigma}_p = 2.2\pm0.4$ km s$^{-1}$ reported in \S 3.5 refers to the {\it intrinsic}
velocity dispersion for Palomar 13: i.e., the contribution from measurement uncertainties has been
removed (see, Pryor \& Meylan 1993).
Our velocity uncertainties have been calculated empirically from 53 distinct 
radial velocities of faint metal-poor stars (accumulated during
multiple observing runs) so we are confident that the measured dispersion
is {\it not} the result of underestimated velocity uncertainties. Such an explanation
would require $\alpha \simeq 20$ km s$^{-1}$ (see Equation~\ref{eq2}), whereas we measure
$\alpha = 9.0$ km s$^{-1}$ and find ${\alpha} \le 12.8$ km s$^{-1}$ at 99\% confidence.

Bright red giants in globular clusters often show evidence for radial velocity variations 
in the range 2-5 km s$^{-1}$ (Gunn \& Griffin 1979; Mayor et~al. 1984). This so-called
velocity ``jitter" is thought to arise from pulsational motions in the atmospheres of
these evolved stars. It is unlikely that our measured velocity dispersion
is the result of such velocity variations since long-term monitoring of red giants in nearby globular
clusters has established that the magnitude of this jitter shows a strong luminosity dependence:
i.e., it is significant at the 1 km s$^{-1}$ level only for stars which are 
located within $\sim$ 1 mag of the RGB tip. Our radial velocity
sample includes no stars this bright. In fact, our most luminous member (ORS-118)
has $M_V \simeq -0.3$ mag --- a full 2.3 mag below the tip of the RGB.

Might our estimate for the velocity dispersion be the inflated by ``dynamically-hard" 
spectroscopic binary stars residing in the cluster? There are reasons to believe that 
the binary star fraction in Palomar 13 may be appreciable. 
First, the CMD of Palomar 13 shows both a sizeable blue straggler population
and a possible ``second sequence" --- features which are usually interpreted as
signatures of unresolved binaries (Mateo et~al. 1990; Romani \& Weinberg 1991).
Second, on purely theoretical grounds, the low stellar density and velocity dispersion
of Palomar 13 (for an assumed ``normal" mass-to-light ratio)
suggest that even wide binaries might have survived disruption
over a Hubble time. For instance, for a central velocity dispersion
of ${\sigma}_{p,0} = 1$ km s$^{-1}$, a pair of equal-mass $0.8M_{\odot}$ stars
would be disrupted by encounters with other members only if separated by
more than $a_{\rm crit} \sim $ 1,400 AU (Hills 1984). Such binaries, if on circular
orbits, would have periods of $P \sim 4\times10^4$ yr and orbital velocities of
$v \sim 1$ km s$^{-1}$: i.e., modest, but potentially important, given the low
dispersion expected for Palomar 13.

To explore the possible effects of binary stars on our measured velocity
dispersion, we have carried out Monte-Carlo simulations similar to those described
in Pryor, Latham \& Hazen (1988) and C\^ot\'e et~al. (1996). In brief, we have generated simulated
radial velocity measurements having the identical precision and temporal spacing
as the actual dataset. For each simulation, we take the intrinsic velocity dispersion 
to that predicted by a King-Model having $\Upsilon_V \simeq 2$, 
and calculate the contribution to the measured velocity dispersion of a randomly
drawn percentage, $f_b$, of unresolved spectroscopic binary stars. Binaries are assumed
to have circular orbits. The orbital periods in years, $P$, and mass ratios, $q = M_2/M_1$, are 
drawn at random over the ranges, $-1.0 \le {\log{P}} \le 4.0$ and $-0.6 \le {\log{q}} \le 0.0$.
For each $f_b$, we generate 1000 artificial datasets and calculate the
mean cluster dispersion in precisely the same manner as the actual for the actual data;
the percentage of binaries is varied until the median dispersion found in 1000 simulated
datasets is equal to the value obtained using the actual observations. 
These simulation suggest that a value of $f_b \simeq 30\%$ produces a median velocity dispersion
which best matches the observed value of ${\sigma}_p = 2.2$ km~s$^{-1}$. If it is
assumed that the binaries do not have circular orbits, but instead have a ``thermal" 
distribution of eccentricites, then the percentage of binaries needed to explain the
measured dispersion would be roughly twice as large (see Pryor et~al. 1988; C\^ot\'e et~al. 1996).

By definition, the velocity dispersion {\it profile} predicted by these simulations differs
from that of the input King-Michie model only in normalization since we
have made the first-order assumption that the binaries have the same density
distribution as the single stars. However, if the binaries are instead assumed to be
more centrally concentrated than the other stars --- perhaps as a result of mass
segregation --- then the expected velocity dispersion profile would show a steeper decline. 
Needless to say, a {\it rising} velocity dispersion profile would prove difficult to reconcile 
with this scenario. We find weak evidence for an outward increase in the observed
velocity dispersion of Palomar 13 but given the small number of measured velocities, this feature clearly
needs to be confirmed by additional observations. As Figure~\ref{fig6} demonstrates, there are many
promising candidates beyond $R \sim 2^{\prime}$ (i.e., the radii of the most distant
members in the current radial velocity sample).
It would be of considerable interest to measure
velocities for these stars in order to determine if they are members of Palomar 13 and, if
so, to see if the velocity dispersion profile rises with increasing radius.

Likewise, it would be desirable to search directly for velocity variations
associated with possible orbital motions of these stars.  Unfortunately, the long periods of 
the putative binaries would render any such search hopeless; a more promising approach
would be to measure $f_b$ by modeling the ``second sequence" measured from
high-resolution {\it HST} imaging (e.g., Rubenstein \& Bailyn 1997). 

Finally, we note that spectroscopic binaries may offer a viable explanation for the 
observed high velocity dispersion, but they cannot on their own explain the anomalous
surface density profile.

\subsection{Disruption?}

An underlying assumption of the dynamical analysis presented in \S 4 is that of virial
equilibrium. In light of the large velocity dispersion measured for Palomar 13, it is
prudent to investigate the validity of this assumption, which has recently been called into
question for many of the dwarf satellites and globular clusters in the Galactic halo 
(Majewski et al. 2001). Simply put, could Palomar 13 be in the process of disruption? 
This possibility was advocated by Siegel et al. (2001), who combined
their proper motion measurements with the systemic radial velocity presented here to derive an orbit for
Palomar 13. They concluded that Palomar 13 is just $\sim$ 70 Myr past perigalacticon ($R_p \simeq 11.2$ kpc)
and suggested that it has undergone catastrophic tidal heating during its recent approach.
Since N-body simulations of tidal tails in globular clusters (Combes, Leon \& Meylan 1999) 
show that the time taken for evaporating stars to diffuse along the cluster orbit can exceed 
this timescale, it is worth examining this possibility further.

Siegel et al. (2001) noted the presence of a population of ``extra-tidal" stars in their
surface density profile. We too find a possible excess in the Palomar 13 surface density profile,
beginning at $R \sim$ 1$^{\prime}$. This feature is evident in both the CFHT
and LRIS surface density profiles, and the CFHT photometry suggests that many of these
stars have magnitudes and colors which are consistent with those expected for Palomar 13
members (see \S 3.4). Thus, we confirm the existence of this feature; the issue is whether 
the stars associated with it are truly evaporating from Palomar 13, or
if they are instead bound members.

As shown in Figure~\ref{fig5}, the surface density profile for Palomar 13 shows a roughly
power-law decline for $R~\gae$~0\farcm3. This behavior is most pronounced 
beyond $R~\sim$~1$^{\prime}$, where there may be some evidence for a ``break" in the profile.
Johnston et~al. (1999; 2001) have carried out N-body simulations of satellites orbiting in the 
Galactic tidal field; they find that the radial distribution of unbound stars in the 
outer regions of disrupting satellites should define a power-law profile --- broadly similar 
to what we observe for Palomar 13. 
The dashed lines in Figure~\ref{fig5} show power-law profiles, 
${\Sigma}$($R$) $\propto$ $R^{-{\gamma}}$, having representative exponents of 
${\gamma} = 0.5,~1.5~{\rm and}~2.5$. Over the range 1$^{\prime}$ $\lae$ $R \lae$ 4$^{\prime}$, we measure
$\gamma = 1.8\pm0.2$ for Palomar 13. This value is steeper than the formal
value of $\gamma \simeq 1$ estimated by Johnston et~al. (1999) but is probably consistent
with expectations given uncertainties in the numerical treatment (e.g., the initial density
profile, viewing geometry and orbital phase; Johnston et~al. 2001).

If this interpretation is correct, then a crude estimate for the mass loss rate may be made
as follows. Johnston et~al. (1999) present two simple expressions for the fractional
mass loss rate using extra-tidal stars,
\begin{equation}
\begin{array}{lr}
(df/dt)_1 = \cos({\theta}){{{\pi}R_{b}N_{xt}} / [(R_{xt}-R_b)N_{b}P]} \\
(df/dt)_2 = \cos({\theta}){{\Sigma}(R_b)2{\pi}R_b^2/ [N_bP]}, \\
\end{array}
\label{eq8}
\end{equation}
where $\theta$ is the angle subtended by our line of sight and the plane perpendicular to
the satellite's motion, $R_b$ in the ``break" radius where the surface density excess begins,
$R_{xt}$ is the radius where the profile of the extra-tidal component
ceases to be well defined, $N_{xt}$ is the number of stars in the range $R_b \le R \le R_{xt}$,
and $P$ is the satellite's {\it azimuthal} period (taken to be 2.0 Gyr following Dinescu et~al. 2001). 
Using the space velocity of Siegel et~al. (2001) and the surface density profile shown in 
Figure~\ref{fig5}, we find $(df/dt)_1 \sim 0.2$ Gyr$^{-1}$ and
$(df/dt)_2 \sim 0.15$ Gyr$^{-1}$. There are significant uncertainties involved in these
calculations but it is clear that, if this interpretation of the surface density profile is
correct, then both estimators point to enormous recent mass loss. The inferred mass loss rate is 
far higher than those estimated by Leon, Meylan \& Combes (2000) from star counts in a sample of 
20 Galactic globular clusters --- e.g., in the case of $\omega$ Centauri, Leon et~al. (2000) find that 
only $\approx$ 1\% of the cluster's mass has been lost during the last tidal encounter. Taken at 
face value, the mass loss rates for Palomar 13 suggest that it has suffered severe heating, 
and will disrupt entirely within the next few Gyrs. 

On the other hand, there may be some difficulties with this interpretation. N-body simulations of
the tidal heating of satellites indicate that the tidally stripped material tends to diffuse along 
the orbits, forming leading and trailing streams. We have examined our star counts 
for evidence of such streams by measuring the mean surface density of stars in four quadrants 
over the radial range $1^{\prime} \le R \le$ 2\farcm88. The inner limit corresponds to the the 
break radius, $R_b$, while the outer limit is the most distant radius at which our geometric 
coverage is still 100\% complete. The results of this exercise are shown in Figure~\ref{fig9}. 
We find only weak evidence that these stars are oriented along a preferred direction. The density 
maxima of the best-fit sinusoid shown in this figure are found at position angles of ${\Theta}_p = 
15\pm11^{\circ}$ and $195\pm11^{\circ}$. The agreement between this orientation and those of
the absolute proper motion vector, ${\Theta}_p = 48^{\circ}$, and the Galactic center direction,
${\Theta}_p = 240^{\circ}$, is unexceptional: i.e., the differences in position
angles are ${\Delta}{\Theta}_p = 33\pm11^{\circ}$ and $45\pm11^{\circ}$, respectively. There
is poor agreement with the vector perpendicular to the Galactic disk, which has 
${\Theta}_p = 327^{\circ}$. If the peculiar surface density
profile of Palomar 13 is indeed the result of tidal heating, then this lack of a preferred orientation 
for the putative extra-tidal stars is somewhat puzzling. It would be worthwhile investigating this issue
using star counts which extend over a wider field, preferably well beyond the modest range
radial range ($R \lae 3^\prime$) examined above.

Piatek \& Pryor (1995) have used N-body simulations to show that it is possible to heat the outer 
regions of satellites during close passages with the Galaxy. Such encounters often have the
effect of inducing {\it ordered} motions at large radii while leaving the central velocity dispersion 
unchanged. According to the simulations, the gradient in velocity is usually maximized along the 
vector which points to the Galactic center. This vector is shown by the long arrow in Figure~\ref{fig1}; 
for comparison, the short arrow shows the direction of the cluster motion as measured by Siegel et~al.
(2001), while the intermediate arrow denotes the direction perpendicular to the Galactic plane. As
shown in Figure~\ref{fig10}, we find no statistically significant evidence for ``rotation" among our 
program stars, as might be expected for stars in the leading or trailing streams of a disrupting 
cluster (see also Figure~\ref{fig1} where the velocity residuals are illustrated on a star-by-star 
basis).

Finally, we note that there is an additional (albeit indirect) constraint on the mass and 
dynamical status of Palomar 13. The tidal radius of a satellite object orbiting in the 
Galactic potential is given by
\begin{equation}
r_t = R_p[M/M_G(3+e)]^{1/3}
\label{eq9}
\end{equation}
where $R_p$ is the distance from the Galactic center at perigalacticon, $M_G$
is the interior Galactic mass at this point, $e$ is the orbital eccentricity of the
satellite, and $M$ is its mass (King 1962). If we assume a canonical mass-to-light for Palomar 13
(e.g., ${\Upsilon}_V = 2$), we may use equation (9) to obtain a crude estimate for the expected
tidal radius. Adopting $e = 0.76$ and $R_p = 11.2$ kpc for Palomar 13 (Siegel et~al. 2001), and 
assuming using $M_G(R_p) = {{\Theta}_c}^2R_p/G$
(Fich \& Tremaine 1991) and ${\Theta}_c = 220$ km s$^{-1}$, we find $r_t \sim 26$ pc for
a total luminosity of $L_V = 3\times10^3$~$L_{V,{\odot}}$. This is remarkably close to the tidal radius of 
$r_t = 23$ pc for the King-Michie model of Siegel et al. (2001), which is shown as the dotted curves in 
Figures~\ref{fig5} and \ref{fig7}. This is consistent
with the view that the excess in the surface density profile is a population of extra-tidal stars. 
The alternative interpretation --- that the much larger tidal radius implied 
by the best-fit King-Michie model in Figure~\ref{fig7} is evidence for dark matter in Palomar 13 ---
will be examined in \S 5.4

\subsection{Baryonic Dark Matter: Low-Mass Stars and/or Heavy Remnants?}

Taken at face value, the measured velocity dispersion for Palomar 13 implies
a total mass of $M \sim (1.1^{+0.4}_{-0.3})\times10^5~M_{\odot}$, far greater than that
expected on the basis of the observed luminosity: i.e., $M  \equiv {\Upsilon}_V{L_V} \sim 6\times10^3~M_{\odot}$.
The obvious baryonic candidates for this additional mass would be low-mass main-sequence
stars, or heavy stellar remnants such as white dwarfs and neutron stars. Both varieties of heavy
remnants are known to populate globular clusters (see Hut et al. 1992) and
the contribution of white dwarfs to the overall mass budget is thought to increase as 
clusters evaporate (e.g., Vesperini \& Heggie 1997). In this case, the low-luminosity of 
Palomar 13 might be interpreted as indirect evidence for an advanced stage of dynamical 
evolution. However, if one accepts the view that Palomar 13 is in dynamical equilibrium 
(see above), then this would require $\sim$ 95\% of the total mass to reside in heavy remnants. 
This percentage is {\it far} higher than that found in nearby, well-studied clusters: e.g., 
Meylan (1987; 1989) finds the percentage of heavy remnants (mainly white dwarfs) in 47 Tucanae 
and $\omega$ Centauri to be $20\pm5$\% and $25\pm15$\%, respectively (though the
dynamical mass found here for Palomar 13 is only half that contained in
heavy remnants in 47 Tuc, by virue of the latter's much larger mass).
Unfortunately, a direct search for white dwarfs in Palomar 13 would be 
extremely challenging since even the brightest such objects are
expected to have $V \gae 27$ mag at this distance (Richer 
et ~al. 1997)\altaffilmark{9}\altaffiltext{9}{Catalogs of
pulsars and X-ray sources associated with globular clusters list no such objects 
in Palomar 13 (Kulkarni \& Anderson 1996; Verbunt 2001).}

On the other hand, a determination of the main-sequence luminosity function down to this level 
(corresponding to $M_V = 10$ mag and ${\Upsilon}_V \simeq$ 25; Silvestri et~al. 1998) is well 
within the reach of {\it HST}, and could put interesting constraints on the mass contained 
in low-mass, main-sequence stars. However, as shown in Figure~\ref{fig11}, the available data
suggest that the main-sequence luminosity function of Palomar 13 is adequately described
by a rather {\it flat} mass function, with exponent $x \sim 0.0$ over the range 
$21.5 \lae V \lae 24$ (equivalent to $0.79 \lae M/M_{\odot} \lae 0.61$; Bergbusch \& 
VandenBerg 1992). If the initial mass function of Palomar 13 had a more nearly universal
form (for instance, near the ``Salpeter value" of $x = 1.35$; Salpeter 1955), then the flat mass function
we find may be evidence for advanced dynamical evolution and the preferential evaporation 
of low-mass stars. In short, the available data provide no reason to believe that the mass function 
of Palomar 13 is abnormal in any way, but firm conclusions must await deeper observations

\subsection{Non-Baryonic Dark Matter?}

In \S 5.2, it was found that, for an assumed mass-to-light ratio of ${\Upsilon}_V = 2$ (i.e.,
a value appropriate for typical globular clusters), the tidal radius of Palomar 13 predicted
by equation (9) was in good agreement with the tidal radius of the low-concentration ($c \sim 0.7$) \
King-Michie model adopted by Siegel et al. (2001). As discussed in {\S} 3.5 and \S 4, it is possible to
obtain an excellent match to the observed surface brightness and surface density profiles
using a King-Michie model which has $c = 2.0\pm0.15$. In this case, the tidal radius of
Palomar 13 is $r_t \sim 180$ pc (with a formal uncertainty of $\pm$ 23\% based on our
Monte-Carlo simulations). From equation (9), we see that a tidal radius of this size 
suggests a mass of $M \sim 2\times10^6$~$M_{\odot}$ --- i.e., considerably larger than our best-estimate
for the dynamical mass of Palomar 13 (assuming virial equilibrium) yet still far in excess
of the {\it expected} mass. The actual uncertainties in the fitted tidal radii are difficult to 
gauge, but are unlikely to be smaller than $\sim$ 50\%, which translates directly into a 
nearly a factor of two in mass. Thus, we conclude that the tidal radii inferred from the 
surface brightness and surface density profiles of Palomar 13 are: (1) highly dependent on
how one chooses to model the profile; and (2) consistent with either the tidal disruption
of an otherwise normal globular cluster, or the presence of substantial dark matter component
in this object.

In light of these results, it is natural to ask if the high velocity dispersion of Palomar 13 
might also be the signature of a non-baryonic dark matter halo, similar
to those belonging to Local Group dSph galaxies (Aaronson 1983; Mateo 1998;
$c.f.,$ Klessen \& Kroupa 1998). This possibility is particularly topical given recent claims
from CDM simulations that the halos of large galaxies may contain large numbers of low-mass,
``dark satellites" (e.g., Hirashita et al. 1999; Klypin et al. 1999; Bullock et al. 2000).

To examine this possibility, we consider a simple two-component model. First, we take the 
mass density profile of the luminous stellar component to be given by a Dehnen (1993) model,
\begin{equation}
{\rho}_{*,m}(r) = [(3 - \gamma)/4](\Upsilon\rho^0_{*,l}/{\pi}a^3)(r/a)^{-{\gamma}}(1 + r/a)^{{\gamma} - 4}
\label{eq10}
\end{equation}
where ${\rho}^0_{*,l}$ is the central {\it luminosity} density of this component, $a$ 
is its scale radius, and $\Upsilon$ is its mass-to-light ratio. By contrast, 
the dark halo is assumed to have the form
proposed by Navarro, Frenk \& White (1997):
\begin{equation}
{\rho}_d(r) = \rho^0_{d}(r/r_s)^{-1}(1 + r/r_s)^{-2}
\label{eq11}
\end{equation}
where $r_s$ is the scale radius of the dark halo, and $\rho^0_{d}$ is the central dark matter density.
Once these density distributions are specified, the velocity dispersion profile
follows from the the Jeans equation,
\begin{equation}
{d \over dr}[{\rho}_*(r){\sigma}_r^2(r)] + {2{\beta}{\rho}_*(r){\sigma}_r^2(r) \over r} = -{GM(r) \over r^2}{\rho}_*(r),
\label{eq12}
\end{equation}
where $\sigma_r(r)$ is the radial component of the velocity dispersion and $M(r)$ is the combined 
masses of the luminous and dark components. For our purposes, we assume that the velocity dispersion tensor 
is isotropic (i.e., $\beta \equiv 0$). This theoretical
velocity dispersion profile is then projected onto the plane of the sky, and compared
directly to the observed velocity dispersion. 

Before doing so, we require estimates of $\gamma$ and $a$ from the surface brightness profile. 
The long-dashed curve in Figure~\ref{fig7} shows the best-fit Dehnen model, with $\gamma = 1.46\pm0.06$ 
and $a = 135\pm9^{{\prime}{\prime}}$ (15.9$\pm$1.0 pc), overlaid on the observed 
surface brightness profile. For an assumed mass-to-light ratio of $\Upsilon_V = 2$, the total mass 
associated with this stellar component is $M = 7\times10^3~M_{\odot}$. 
We also assume that, in projection, the dark matter matches that of the 
surface brightness profile, giving $r_s = 24\pm1^{{\prime}{\prime}}$ (2.8$\pm$0.1 pc). This NFW model ---
fitted to the observed surface brightness profile --- is shown by the solid curve in Figure~\ref{fig7}.
The central {\it mass density} of this model is then left as a free parameter which is adjusted until a
reasonable match to the observed velocity dispersion is achieved. 
The corresponding velocity dispersion profile 
for a central mass density of $\rho^0_{d} = 80\pm31~M_{\odot}$ pc$^{-3}$
is shown as the solid curve in Figure~\ref{fig8} (i.e., curve 3). This curve provides an adequate (though by
no means unique) description of the observed velocity dispersion profile including the
possible maximum at $R \sim 1^{\prime}$. Although the mass of an NFW profile 
diverges as $r \rightarrow \infty$, we note that over the range of our radial velocity and surface brightness 
observations, the implied mass-to-light ratio of this two component model is 
${\Upsilon}_V = 19^{+8}_{-7}$ --- about half that found using a King-Michie model,
but still well above that expected for a normal stellar population.

How does the central density of this putative dark halo compare to those found for the dSph 
satellites of the Milky Way? Table 1 of Mateo (1998) lists structural parameters 
for nine Galactic dSph galaxies having measured velocity dispersions: the central densities 
of these galaxies span the range
$0.03 \le \rho^0 \le 0.6~M_{\odot}~{\rm pc}^{-3}$, with a mean value of 
$\langle \rho^0\rangle = 0.26\pm0.07~M_{\odot}~{\rm pc}^{-3}$. Strictly speaking, 
these values refer to the {\it combined} densities of 
the dark and luminous components, $\rho^0 = \rho^0_{d} + \rho^0_{*,m}$. However, in view of
the low stellar densities of these galaxies 
(e.g., ${\rho}^0_{*,m} \sim 0.01~M_{\odot}~{\rm pc}^{-3}$),
$\rho^0 \simeq \rho^0_{d}$ to good approximation. {\it Thus, the dark matter
density needed to explain the measured velocity dispersion of Palomar 13 exceeds
that of typical dSph galaxies by more than two orders of magnitude.} 

Following Kormendy (1990), we show in
Figures~\ref{fig12} and ~\ref{fig13} the scaling relations between central density, 
velocity dispersion, $\sigma$, absolute blue magnitude, $M_B$, core radius, $r_c$, 
and central $B$-band surface brightness, $\mu_B$, for a sample of Local Group dSph galaxies 
(Mateo 1998), dwarf irregular and spiral galaxies (C\^ot\'e, Carignan \& Freeman 2000 and 
references therein) and Galactic globular clusters (Pryor \& Meylan 1993; Harris 1996). 
Note that the densities plotted here
refer to those of the {\it dark halos}, with the exception of the globular clusters; 
for these objects, the densities are those calculated directly from the measured velocity 
dispersions. For Palomar 13, which is indicated by the large filled circle, we take the 
structural parameters for its putative dark matter halo to be those given by the best-fit 
NFW model.
As these figures demonstrate, the central dark matter density
in Palomar 13 is indeed much larger than those found for dwarf galaxies. On the other
hand, it is also consistent with that expected from an extrapolation of galactic 
scaling relations to low luminosities. Indeed, it is remarkable that Palomar 13 falls 
so near the extrapolations of the tight scaling relations for dwarf galaxies, although
this agreement cannot be taken as unambiguous evidence for a common origin since Palomar 13
also falls near the edges of the more heterogeneous distributions for Galactic globular 
clusters (and, in the scaling relations involving $\mu_B$, its location may be more consistent 
with the globular cluster distributions).
Thus, the evidence from the observed scaling relations concerning the proper classification 
of Palomar 13 as a {\it bone~fide} globular cluster or an extreme dark-matter dwarf galaxy, is 
ambiguous. In any event, we note that a naive extrapolation of the dwarf galaxy scaling relations
to globular cluster scales predicts central densities for dark matter halos in
these objects that are 
sufficiently low (i.e., $1 \lae {\rho^0_d} \lae 100~M_{\odot}~{\rm pc}^{-3}$) to make
their detection extremely challenging in all but the faintest clusters. In other words, 
for the vast majority of globular clusters, the inferred dark matter densities would fall well 
below the {\it baryonic} densities. 

Obviously, it would be of considerable interest to measure velocity dispersions for additional
faint, low surface brightness clusters. An especially interesting target is AM 4, a cluster
which shares many of the characteristics of Palomar 13:  i.e., low luminosity 
($L_V \sim 400~L_{V,{\odot}}$), low surface brightness ($\mu_{V,0} \sim 24.75$ mag arcsec$^{-2}$), 
and low metallicity ([Fe/H] $\sim -2.0$ dex). Like Palomar 13,
it is also occupies a relatively isolated location in
the Galactic halo, with $R_G = 25.5$ kpc. The central velocity dispersion predicted by
equation (7) for a ``normal" globular cluster of this surface brightness would be
${\sigma}_{{\rm p},0} \lae 0.3$ km s$^{-1}$, and it would be interesting to see if this
expectation is borne out by radial velocity measurements.

We consider three final scaling relations for dwarf galaxies in Figure~\ref{fig14}.
In the upper panel of this figure, metallicity is plotted against absolute
magnitude for dE/dSph galaxies, along with the best-fit linear relation from
C\^ot\'e et al. (2000). Note that the relationship between metallicity and luminosity
constitutes a key difference between dwarf galaxies and globular clusters: i.e., galaxies 
spanning a range of $\sim 10^5$ in luminosity follow a tight relation between
metallicity and luminosity (e.g., Brodie \& Huchra 1991; Caldwell et al. 1992), whereas 
globular clusters obey no such correlation. While it is by no means clear that a simple
extrapolation of this linear relation to $M_V \sim -4$ is justifiable, Palomar 13 is 
nevertheless considerably more metal-rich than might naively be expected for a dwarf 
of this luminosity. 
Furthermore, its CMD shows no evidence for an abundance spread, unlike Local 
Group dSph galaxies which often show internal dispersions in metallicity of $\sim$ 0.35 dex 
(e.g., C\^ot\'e et al. 2000; Shetrone, C\^ot\'e \& Sargent 2001).

In the middle panel of Figure~\ref{fig14}, mass-to-light ratio is plotted against
absolute magnitude for all Local Group dE/dSph galaxies having measured velocity
dispersions. The dashed curve shows the relation expected
for dwarf consisting of luminous stellar components with ${\Upsilon}_V = 2$
that are embedded in dark matter halos of mass $M = 2\times10^7~M_{\odot}$, as
proposed by Mateo et al. (1993). The mass-to-light ratio predicted for Palomar 13 by
this relation would be ${\Upsilon}_V \sim 7000$, making it an extreme outlier.
On the other hand, Palomar 13 {\it does} seem to follow the correlation 
between mass-to-light ratio, metallicity and central surface brightness 
for Local Group dwarfs, shown in the lower panel of Figure~\ref{fig14} 
(Prada \& Burkert 2001). We can only conclude that the evidence regarding a possible connection
between Palomar 13 and dark-matter-dominated dwarf galaxies is ambiguous 
and that, in light of the complexity of gas cooling, star formation and feedback in low-mass 
systems (see, e.g., Figure~8 of Mateo 1998), a definitive conclusion is not possible
at the present time.

\subsection{Modified Newtonian Dynamics?}

A much different interpetation of the large velocity dispersion of dSph galaxies
has been proposed by Milgrom (1983a), who suggested that at low accelerations, $a \ll a_0$, Newtonian
gravity should be revised to include a repulsive term. If this modified theory of
Newtonian dynamics (i.e., MOND) is correct, then this there would be no need to
invoke dark matter as an explanation for the high velocity dispersions of dwarf galaxies
(Milgrom 1983b; 1995). But if MOND is to be a viable alternative to dark
matter, then it must apply to all systems in which the the acceleration falls below
the characteristic MOND value, $a_{\rm 0} \simeq 1.2\times10^{-8}$ cm s$^{-2}$ ---
value chosen to match the rotation curves of low-surface
brightness galaxies (Begeman, Broeils \& Sanders 1991; McGaugh \& de Blok 1998).
Since this condition is met for Palomar 13, we now examine its internal dynamics
within the MOND framework.

There are two relevant accelerations for Palomar 13: the acceleration arising
from the internal mass distribution, and an external component due to the Milky Way.
We approximate the internal acceleration as
$a_{\rm int} = GM(r)/r^2$, where $M(r)$ is mass enclosed with a radius, $r$, for
the best-fit King model shown in Figure~\ref{fig7}. For an assumed
mass-to-light ratio of ${\Upsilon}_V = 2$, the internal acceleration
reaches a maximum value of $a_{\rm int} \simeq 1\times10^{-9}$ cm s$^{-1}$ at
$r \simeq r_c$. The external acceleration may be approximated
as $a_{\rm ext} = {\Theta}_c^2/R_G$, or
$a_{\rm ext} = 6\times10^{-9}$ cm s$^{-1}$ for ${\Theta}_c = 220$ km s$^{-1}$ and
$R_G = 25.3$ kpc. Thus, according to MOND, Palomar 13 is in the 
``quasi-Newtonian weak-field limit" since $a_{\rm int} < a_{\rm ext} < a_{\rm 0}$.
In this regime, the Newtonian mass is related to the MOND mass by a factor,
$\mu$, which falls in the range $1 \lae \mu \lae a_{\rm 0}/a_{\rm ext}$ (Gerhard \& Spergel 1992).
From Table 4, the best-fit King-Michie model has ${\Upsilon}_V \simeq 40^{+24}_{-17}$ in solar units.
Thus, the mass-to-light ratio in MOND would be reduced by --- at most --- a factor of $a_{\rm 0}/a_{\rm ext}$,
giving ${\Upsilon}_V \simeq 20^{+12}_{-9}$. Apparently, the high mass-to-light ratio of
Palomar 13 persists (at the 2$\sigma$ level) even within the MOND framework. We
conclude that this explanation appears less appealing than those involving
tidal disruption or dark matter.

\section{Summary and Conclusions}

An imaging and spectroscopic survey of globular clusters in the outer halo of the Milky 
Way has shown Palomar 13 to exhibit several puzzling features. 

From an abundance analysis 
of our HIRES spectra for a single RGB star and isochrone fitting to the CMD, we 
estimate the metallicity of Palomar 13 to be [Fe/H] = $-1.9\pm0.2$ dex. We find a true distance
modulus of $(m-M)_0 = 16.93\pm0.1$ mag and a distance of $R_S = 24.3^{+1.2}_{-1.1}$ kpc,
placing Palomar 13 squarely in the outer halo. With $M_V = -3.8$ mag, this is the lowest
luminosity object in our survey. The sparsely populated red giant and horizontal 
branches preclude a precise age determination, but a 14 Gyr isochrone is found to provide a reasonable 
match to the observed fiducial sequences. Thus, in terms of its stellar populations,
Palomar 13 appears typical of globular clusters in the outer halo.

However, from a sample of 21 probable members with radial velocity measurements from 
HIRES, we find the systemic velocity and intrinsic velocity dispersion of Palomar 13 to be 
$\langle v_r\rangle_s = 24.1\pm0.5$ km s$^{-1}$ and ${\sigma}_p = 2.2\pm0.4$ km s$^{-1}$, respectively. 
This velocity dispersion is roughly four times higher than that expected on the basis of the observed 
luminosity and concentration. Taken at face value, this dispersion suggests a mass-to-light ratio of 
${\Upsilon}_V = 40^{+24}_{-17}$, with the exact value depending on the details of the modeling.
This mass-to-light ratio is far higher than those found for globular clusters, but
is comparable to those of Galactic dSph galaxies.
The surface density profile is also found to be anomalous among Galactic globular 
clusters, in that it shows a significant number of apparent member stars beyond the 
tidal radius of a low concentration King-Michie model (i.e., $c \sim 0.7$, as
has usually been claimed for Palomar 13). On the other hand, an excellent fit to the 
surface brightness profile over all radii is provided by a King-Michie model with 
$c \sim 2$, as well as by models of the type suggested by Dehnen (1993) and Navarro et al. (1997).

Based on the measured velocity
dispersion and surface density profile, two possibilities present themselves: (1)
either Palomar 13 has recently undergone catastrophic heating and mass loss, as might
be expected on the basis of its orbital properties; or (2) its spatial extent, central 
concentration, and mass have been greatly underestimated in previous 
studies. Other factors, such as a large fraction of spectroscopic binary stars or the
presence of a large number of heavy remnants, may play contribute to Palomar 13's
unusual properties but are, on their own, unlikely to provide a complete explanation
of the observed structural and kinematic properties.
Additional observations --- particularly deep imaging with {\it HST} and ground-based
mosaic cameras, and radial velocity measurements for an expanded sample of stars at 
large radii --- should help discriminate between the various possibilities.
In any event, it is clear that Palomar 13 merits further study, as it offers
new insights into the structure and formation of the Galactic halo.

\acknowledgments

We thank Tad Pryor and an anonymous referee for numerous helpful comments.
We also thank Gilles Bergond for his assistance with the CFHT observations, and Dean McLaughlin
for providing some of the software used in the analysis. PC gratefully
acknowledges support provided by the Sherman M. Fairchild Foundation during the initial
stages of this work.  SJG acknowledges partial support from the Bressler Foundation.
This work was based in part on observations obtained at the W.M. Keck 
Observatory, which is operated jointly by the California Institute of Technology and 
the University of California. We are grateful to the W.M. Keck Foundation 
for their vision and generosity.

\clearpage

\begin{deluxetable}{lrcc}
\tablecolumns{4}
\tablewidth{0pc}
\tablecaption{Observing Log\label{tab1}}
\tablehead{
\colhead{Telescope} &
\colhead{Date} &
\colhead{Spectrograph} &
\colhead{Imager}
}
\startdata
Keck II & ~~~10/09/1999 &      ...          & LRIS + $BVI$ filters   \\
CFHT    & ~~~14/07/1999 &      ...          & CFH12K + $VR$ filters \\
Keck ~I & 21-22/08/1998 & HIRES + C1 Decker &      ...            \\
Keck ~I & 15-16/10/1998 & HIRES + C5 Decker &      ...            \\
Keck ~I & 15-16/07/1999 & HIRES + C1 Decker &      ...            \\
Keck ~I & ~~~11/08/1999 & HIRES + C1 Decker &      ...            \\
\enddata
\end{deluxetable}

\clearpage

\begin{deluxetable}{lrccccrrrcc}
\tablecolumns{11}
\tablewidth{0pc}                        
\tablecaption{Radial Velocities for Candidate Red Giants in Palomar 13\label{tab2}}                    
\tablehead{                        
\colhead{ID} &                       
\colhead{$R$} &                      
\colhead{$V$} &                       
\colhead{$(B-V)$} &                       
\colhead{$T$} &
\colhead{HJD} &                      
\colhead{$R_{TD}$} &                       
\colhead{$v_r$} &
\colhead{$\langle v_r \rangle$} &
\colhead{P($\mu$)} &
\colhead{Member?} \\
\colhead{} &                      
\colhead{($^{\prime\prime}$)} &
\colhead{(mag)} &
\colhead{(mag)} &     
\colhead{(sec)} &
\colhead{2,450,000+} &                      
\colhead{} &                       
\colhead{(km s$^{-1}$)} &
\colhead{(km s$^{-1}$)} &
\colhead{(\%)} &
\colhead{}
}                        
\startdata                        
ORS-13    & 113 & 16.55 &  0.89 &\p180 &  1046.9063 &  17.61 &   -38.38$\pm$0.48 & -38.90$\pm$0.29 &~0 & N \\
          &     &       &       &\p240 &  1047.9159 &  24.72 &   -39.17$\pm$0.35 &                 &   &   \\
ORS-118   & 102 & 17.00 &  0.93 &\p270 &  1046.9147 &  13.04 &    25.43$\pm$0.64 &  24.92$\pm$0.21 &~0 & Y \\
          &     &       &       &\p600 &  1374.9741 &  27.43 &    25.07$\pm$0.32 &                 &   &   \\
          &     &       &       &\p720 &  1375.9423 &  29.27 &    24.68$\pm$0.30 &                 &   &   \\
ORS-6     & 125 & 17.03 &  1.10 &\p240 &  1046.9104 &  18.00 &     0.68$\pm$0.47 &   0.68$\pm$0.47 &~0 & N \\
ORS-103   & 134 & 17.04 &  1.00 &\p360 &  1046.9196 &  24.48 &   -56.72$\pm$0.35 & -56.72$\pm$0.35 &~0 & N \\
ORS-110   & 100 & 17.21 &  1.03 &\p600 &  1046.9266 &  23.96 &     5.40$\pm$0.36 &   5.40$\pm$0.36 &~0 & N \\
ORS-72    &  60 & 17.64 &  0.85 &\p240 &  1101.7681 &  10.22 &    28.64$\pm$0.80 &  28.79$\pm$0.27 &88 & Y \\
          &     &       &       &\p750 &  1046.9350 &  13.45 &    29.30$\pm$0.62 &                 &   &   \\
          &     &       &       &\p750 &  1047.9235 &  13.14 &    30.05$\pm$0.64 &                 &   &   \\
          &     &       &       &\p900 &  1374.9973 &  24.38 &    28.26$\pm$0.36 &                 &   &   \\
ORS-31    &  95 & 17.76 &  0.83 &\p750 &  1046.9456 &  10.70 &    26.31$\pm$0.77 &  25.09$\pm$0.35 &79 & Y \\
          &     &       &       &\p900 &  1374.9844 &  22.02 &    24.78$\pm$0.39 &                 &   &   \\
ORS-91    &  27 & 17.81 &  0.67 &\p750 &  1046.9554 &   6.72 &    24.46$\pm$1.17 &  24.59$\pm$0.60 &99 & Y \\
          &     &       &       &\p900 &  1375.0091 &  12.39 &    24.64$\pm$0.67 &                 &   &   \\
V2        &   0 & 17.91 &  0.49 & 1200 &  1375.1025 &   9.32 &    25.39$\pm$0.87 &  25.39$\pm$0.87 &...& Y \\
ORS-32    &  62 & 18.02 &  0.89 &\p900 &  1046.9664 &  11.56 &    20.46$\pm$0.72 &  19.68$\pm$0.26 &35 & Y \\
          &     &       &       &\p900 &  1375.1170 &  18.95 &    19.31$\pm$0.45 &                 &   &   \\
          &     &       &       & 1200 &  1376.1122 &  18.60 &    18.69$\pm$0.46 &                 &   &   \\
          &     &       &       & 1200 &  1401.9353 &  17.23 &    20.90$\pm$0.49 &                 &   &   \\
ORS-41    &  64 & 18.59 &  0.76 &\p320 &  1101.7397 &   3.91 &    18.37$\pm$1.84 &  19.24$\pm$0.47 &88 & Y \\
          &     &       &       & 1200 &  1046.9987 &   8.72 &    18.91$\pm$0.93 &                 &   &   \\
          &     &       &       & 1500 &  1375.9567 &  15.56 &    19.43$\pm$0.54 &                 &   &   \\
ORS-1     & 151 & 18.59 &  0.82 & 1200 &  1046.9799 &  13.35 &   -37.48$\pm$0.63 & -37.48$\pm$0.63 &~0 & N \\
ORS-101   &  39 & 18.65 &  0.94 & 1200 &  1047.0141 &  18.88 &   -17.39$\pm$0.45 & -17.39$\pm$0.45 &~0 & N \\
ORS-14    &  92 & 18.76 &  0.74 & 1500 &  1047.0337 &  17.63 &   -55.31$\pm$0.48 & -55.31$\pm$0.48 &~0 & N \\
ORS-86    &  14 & 18.81 &  0.77 & 1800 &  1047.8735 &  10.01 &    24.47$\pm$0.82 &  24.47$\pm$0.82 &99 & Y \\
ORS-36    &  41 & 18.98 &  0.75 & 1800 &  1047.0760 &   9.13 &    25.29$\pm$0.89 &  25.29$\pm$0.89 &93 & Y \\
ORS-63    &   6 & 19.02 &  0.76 & 1800 &  1047.0530 &   9.12 &    26.09$\pm$0.89 &  25.09$\pm$0.42 &62 & Y \\
          &     &       &       & 2100 &  1375.0805 &  18.35 &    24.82$\pm$0.47 &                 &   &   \\
ORS-87    &  16 & 19.05 &  0.72 & 1800 &  1047.9021 &   9.06 &    26.55$\pm$0.90 &  26.55$\pm$0.90 &94 & Y \\
910       & 111 & 19.28 &  0.73 &\p600 &  1101.7614 &   4.15 &    23.77$\pm$1.75 &  23.77$\pm$1.75 &89 & Y \\
911       &  52 & 19.36 &  0.75 &\p600 &  1101.7306 &   3.86 &    23.19$\pm$1.86 &  23.19$\pm$1.86 &99 & Y \\
ORS-23    &  77 & 19.49 &  1.50 & 2400 &  1047.1068 &  10.01 &    18.57$\pm$0.82 &  18.57$\pm$0.82 &~0 & N \\
915       &   9 & 19.58 &  0.73 & 1800 &  1375.0284 &  11.68 &    24.69$\pm$0.71 &  24.69$\pm$0.71 &...& Y \\
ORS-18    & 112 & 19.62 &  0.73 & 1000 &  1102.7253 &   4.36 &    25.16$\pm$1.68 &  25.16$\pm$1.68 &98 & Y \\
ORS-38    &  40 & 19.65 &  0.74 & 1000 &  1101.7489 &   4.49 &    19.52$\pm$1.64 &  21.64$\pm$0.72 &88 & Y \\
          &     &       &       & 2400 &  1376.0856 &  10.57 &    22.11$\pm$0.78 &                 &   &   \\
931       &  18 & 19.70 &  0.72 & 2100 &  1375.0528 &  11.70 &    25.45$\pm$0.71 &  25.45$\pm$0.71 &91 & Y \\
ORS-78    &  19 & 19.73 &  0.72 & 1000 &  1102.7404 &   5.26 &    22.34$\pm$1.44 &  22.34$\pm$1.44 &78 & Y \\
ORS-50    &  22 & 19.76 &  0.72 & 2700 &  1376.0226 &   9.66 &    24.15$\pm$0.85 &  24.15$\pm$0.85 &98 & Y \\
ORS-96    &  32 & 19.80 &  0.71 & 1000 &  1102.7542 &   6.52 &    25.38$\pm$1.20 &  25.38$\pm$1.20 &99 & Y \\
ORS-5     & 117 & 19.86 &  0.72 & 2700 &  1375.9867 &   9.99 &    20.95$\pm$0.82 &  20.95$\pm$0.82 &98 & Y \\
ORS-88    &  37 & 20.11 &  0.71 & 2100 &  1376.0541 &   7.34 &    24.61$\pm$1.08 &  24.61$\pm$1.08 &78 & Y \\
\enddata                        

\end{deluxetable}                        

\clearpage

\begin{deluxetable}{lcccc}
\tablecolumns{5}
\tablewidth{0pc}
\tablecaption{Observed and Derived Parameters for Palomar 13\label{tab3}}
\tablehead{
\colhead{Parameter} &
\colhead{Symbol} &
\colhead{Value} &
\colhead{Units} &
\colhead{Reference\tablenotemark{a}}
}
\startdata
Right Ascension           & $\alpha$(J2000) & 23:06:44.8 & h:m:s & 1\\
Declination               & $\delta$(J2000) & +12:46:18  & $\circ$:$\prime$:$\prime\prime$ & 1\\
Apparent Distance Modulus & $(m-M)_V$ &17.27$\pm$0.1  & mag & 2\\
True Distance Modulus     & $(m-M)_0$ &16.93$\pm$0.1  & mag & 2\\
Reddening                 & $E(B-V)$  &0.11$\pm$0.02 & mag & 3\\
Galactic Coordinates      & $(l,b)$   &$(87.1,-42.7)$ & deg & 1\\
Distance                  & $R_S$     & 24.3$^{+1.2}_{-1.1}$  & kpc & 2\\
Galactocentric Distance   & $R_G$     & 25.3$^{+1.2}_{-1.1}$  & kpc & 2\\
Metallicity               & [Fe/H]    &-$1.9\pm0.2$ & dex & 2\\
Perigalacticon distance   & $R_p$     & 11.2 & kpc & 1\\
Apogalacticon distance    & $R_a$     & 80.2 & kpc & 1\\
Orbital Period            & $P$       &  1.1 & Gyr & 1\\
Eccentricity              & $e$       &  0.76 &    & 1\\
Maximum Likelihood Mean Velocity       & $\langle v_r\rangle_s$ & $24.1\pm0.5$ & km s$^{-1}$& 2\\
Maximum Likelihood Dispersion          & ${\sigma}_p$ & $2.2\pm0.4$ & km s$^{-1}$& 2\\
\multicolumn{5}{c}{{\it {\underline{King-Michie Model of Siegel et al. (2001)}}}}\nl
Core Radius               & $r_c$     &39                & arcsec & 1\\
                          &           &4.6               & pc     & 1\\
Tidal Radius              & $r_t$     &188               & arcsec & 1\\
                          &           &23                & pc     & 1\\
Central Projected Luminosity Density& $\mu_{V,0}$&23.39             & mag arcsec$^{-2}$ & 2\\
                          &           &15                & $L_{V,{\odot}}$ pc$^{-2}$ & 2\\
Concentration             & $c$       &0.7              &        & 1\\
Integrated Luminosity     & $L_{V}$   &$1.2\times10^3$&$L_{V,{\odot}}$& 1\\
\multicolumn{5}{c}{{\it {\underline{Best-fit King-Michie Model}}}}\nl
Core Radius               & $r_c$     &14$\pm$2          & arcsec & 2\\
                          &           &1.70$\pm$0.24     & pc     & 2\\
Tidal Radius              & $r_t$     &26$\pm$6          & arcmin & 2\\
                          &           &182$\pm$41        & pc     & 2\\
Central Projected Luminosity Density& $\mu_{V,0}$&22.54$^{+0.20}_{-0.17}$     & mag arcsec$^{-2}$ & 2\\
                          &           &33.8$\pm$5.8      & $L_{V,{\odot}}$ pc$^{-2}$ & 2\\
Concentration             & $c$       &2.0$\pm0.15$      &        & 2\\
Integrated Luminosity     & $L_{V}$   &$(2.8\pm0.4)\times10^3$&$L_{V,{\odot}}$& 2\\
Mass-to-Light Ratio       & ${\Upsilon}_V$   &40$^{+24}_{-17}$  & $M_{\odot}/L_{V,{\odot}}$ & 2\\
\multicolumn{5}{c}{{\it {\underline{Dehnen Model}}}}\nl
Power-Law Index           & $\gamma$  &1.46$\pm0.06$     &        & 2\\
Scale Radius              & $a$       &135$\pm$9         & arcsec & 2\\
                          &           &15.9$\pm1.0$      & pc     & 2\\
Integrated Luminosity     & $L_{V}$   &$(3.5\pm0.1)\times10^3$&$L_{V,{\odot}}$& 2\\
\multicolumn{5}{c}{{\it {\underline{NFW Model}}}}\nl
Scale Radius              & $r_s$     &24$\pm$1          & arcsec & 2\\
                          &           &2.8$\pm$0.1       & pc     & 2\\
Central Luminosity Density& $\rho_V^0$&4.3$\pm$0.4     & $L_{V,{\odot}}$ pc$^{-3}$ & 2\\
Central Mass Density      & $\rho_d^0$&80$\pm$31       & $M_{\odot}$ pc$^{-3}$ & 2\\
Central Mass-to-Light Ratio& ${\Upsilon}_V$   &19$^{+8}_{-7}$    & $M_{\odot}/L_{V,{\odot}}$ & 2\\
\enddata
\tablenotetext{a}{References for Table~\ref{tab3}:
(1) Siegel et al. (2001);
(2) This paper;
(3) Schlegel, Finkbeiner \& Davis (1998)}
\end{deluxetable}

\clearpage

\begin{figure}
\plotone{fig01.eps}
\caption{$V$-band image of Palomar 13 taken with the Low-Resolution Imaging Spectrometer
on the Keck II telescope. This image measures 5\farcm8$\times$7\farcm3. The solid circle
shows the core radius ($r_c = 14^{{\prime}{\prime}}$) of the best-fit King-Michie model.
The dashed circle denotes the ``break" radius where the possible excess in the
background-subtracted surface density profile begins: $R_b \simeq 1^\prime$ (see \S 5.1).
The shortest of the three arrows shows the direction of the absolute proper motion from 
Siegel et~al. (2001). The intermediate arrow points in the direction perpendicular to the 
Galactic plane, while the long arrow points in the direction of the Galactic center.
Stars judged to be probable cluster members based on their radial velocities and proper
motions are circled. The size of each circle is proportional to the absolute value of
the difference in the measured radial velocity from the systemic velocity of the
cluster. Red and blue circles denote stars with positive and negative velocity 
residuals, respectively. Non-members are indicated by open squares.
\label{fig1}}
\end{figure}

\clearpage 

\begin{figure}
\plotone{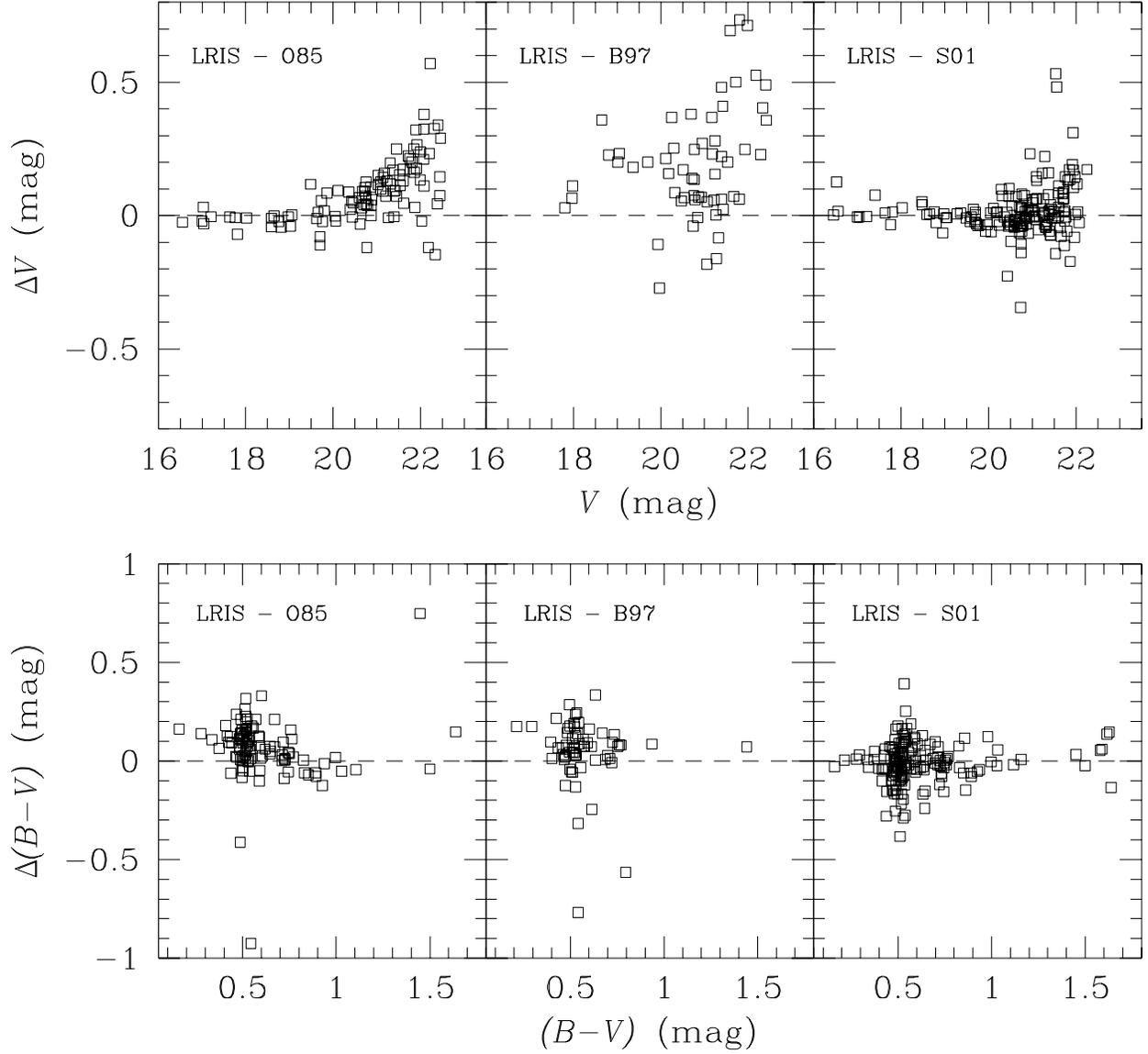}
\caption{(Upper Panels) Difference in $V$ magnitudes for stars in common between our 
LRIS photometric study and those of Ortolani, Rosino \& Sandage (1985; O85),
Borissova, Markov \& Spassova (1997; B97) and Siegel et~al. (2001; S01).
(Lower Panels) Difference in $(B-V)$ color for stars in common between our
LRIS photometric study and those of Ortolani, Rosino \& Sandage (1985; O85),
Borissova, Markov \& Spassova (1997; B97) and Siegel et~al. (2001; S01).
\label{fig2}}
\end{figure}

\clearpage

\begin{figure}
\plotone{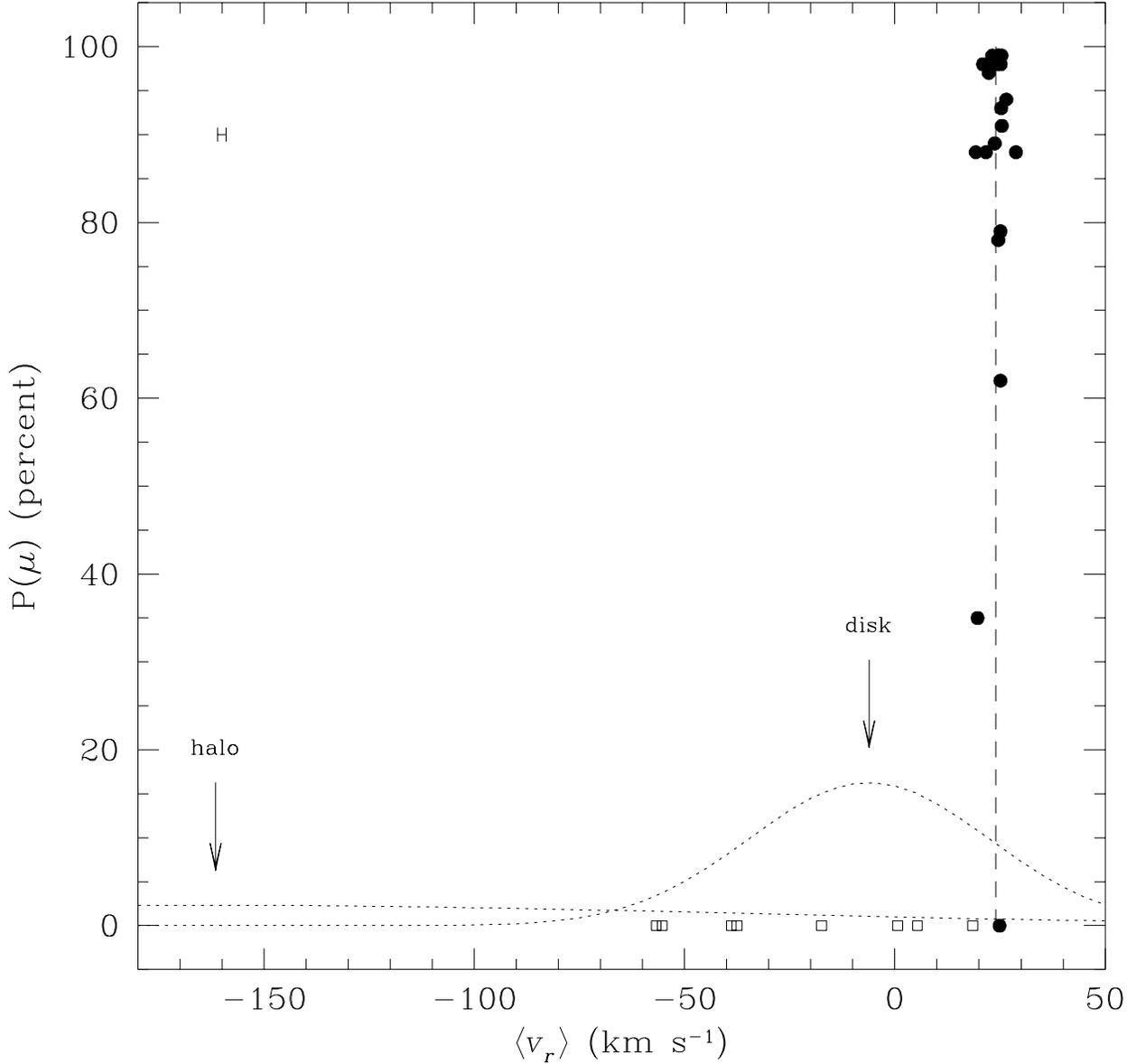}
\caption{Proper motion membership probability from Siegel et~al. (2001), plotted
against mean radial velocity for candidate Palomar 13 stars. The dotted curves
show approximate velocity distributions along this line of sight for disk and halo 
field stars (see text for details). The dashed vertical line indicates the systemic 
velocity of Palomar 13. The error bar in the upper left corner shows the typical
uncertainty in the measured radial velocity.
Open squares show non-members, while probable cluster members are indicated by
the filled circles. Note that this latter sample includes one star (ORS-118) 
which has P(${\mu}$) = 0\% according to Siegel et~al. (2001). We argue that
this star is in fact a cluster member since it: (1) has a radial velocity which
is indistinguishable from cluster systemic velocity; (2) is located precisely on 
the red giant branch in the cluster color-magnitude diagram; and (3) has a 
metallicity of [Fe/H] = $-1.98\pm0.31$ dex measured from our HIRES spectra.
\label{fig3}}
\end{figure}

\clearpage

\begin{figure}
\plotone{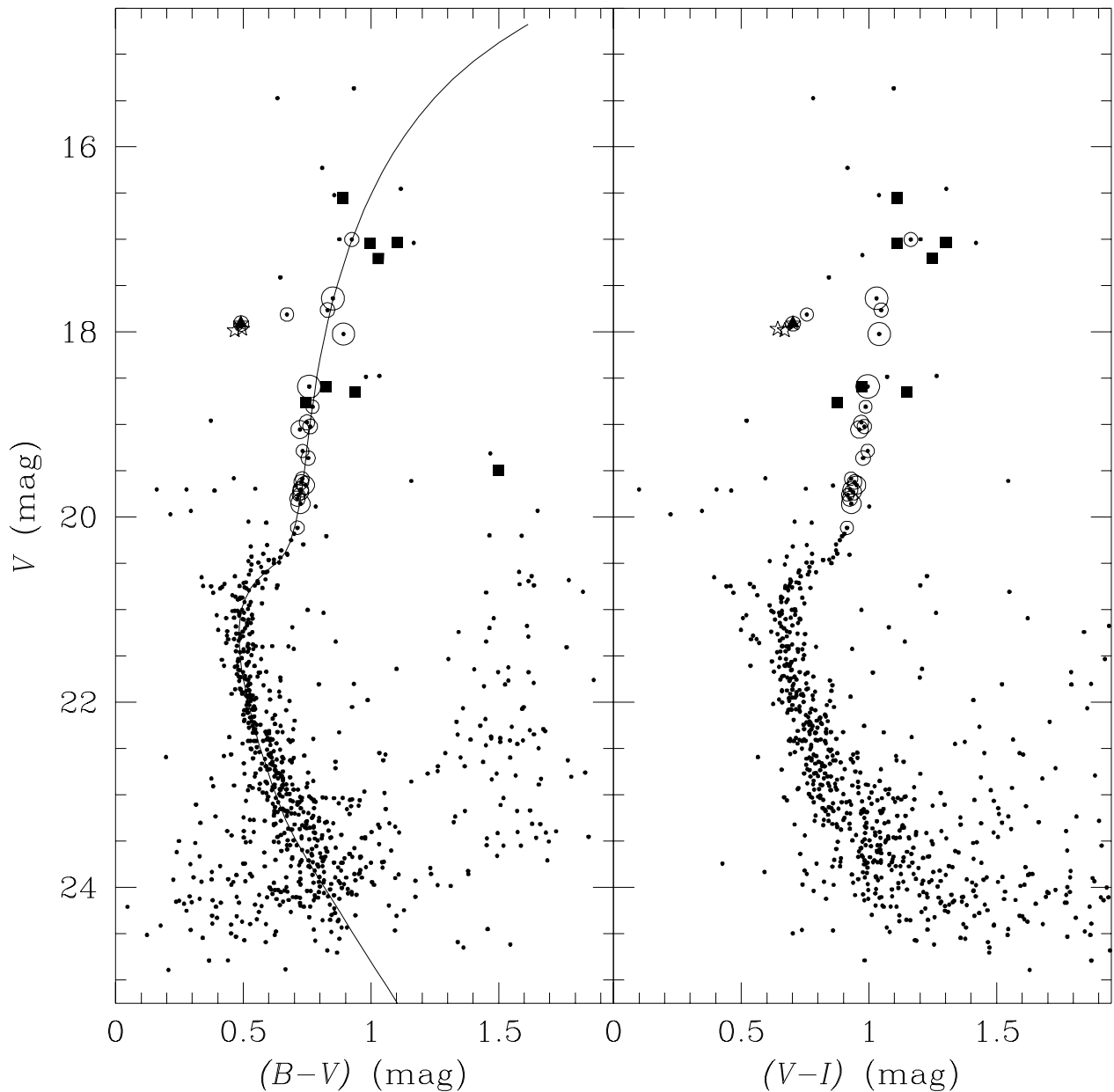}
\caption{(Left Panel) $V$, $(B-V)$ color magnitude diagram for Palomar 13 based on our LRIS images.
A total of 840 objects are plotted in this figure. Open circles indicate 
probable cluster members, determined from our HIRES radial velocities.
The size of each circle is proportional to the absolute value of velocity 
residual with respect to the cluster's systemic velocity. Radial velocity non-member stars are
indicated by the filled squares. Known RR Lyrae variables are indicated by the
stars; the circled variable is V2. The solid curve shows an isochrone from Bergbusch \& 
VandenBerg (1992) having an age of T = 14 Gyr and a metallicity [Fe/H] = $-1.78$, which has
been shifted by $E(B-V)$ = 0.11 mag (Schlegel et~al. 1998) and (m-M)$_{\rm V}$ = 17.27 mag.
(Right Panel) $V$, $(V-I)$ color-magnitude diagram for Palomar 13 derived from our LRIS images.
The symbols are the same as in the previous panel.
\label{fig4}}
\end{figure}

\begin{figure}
\plotone{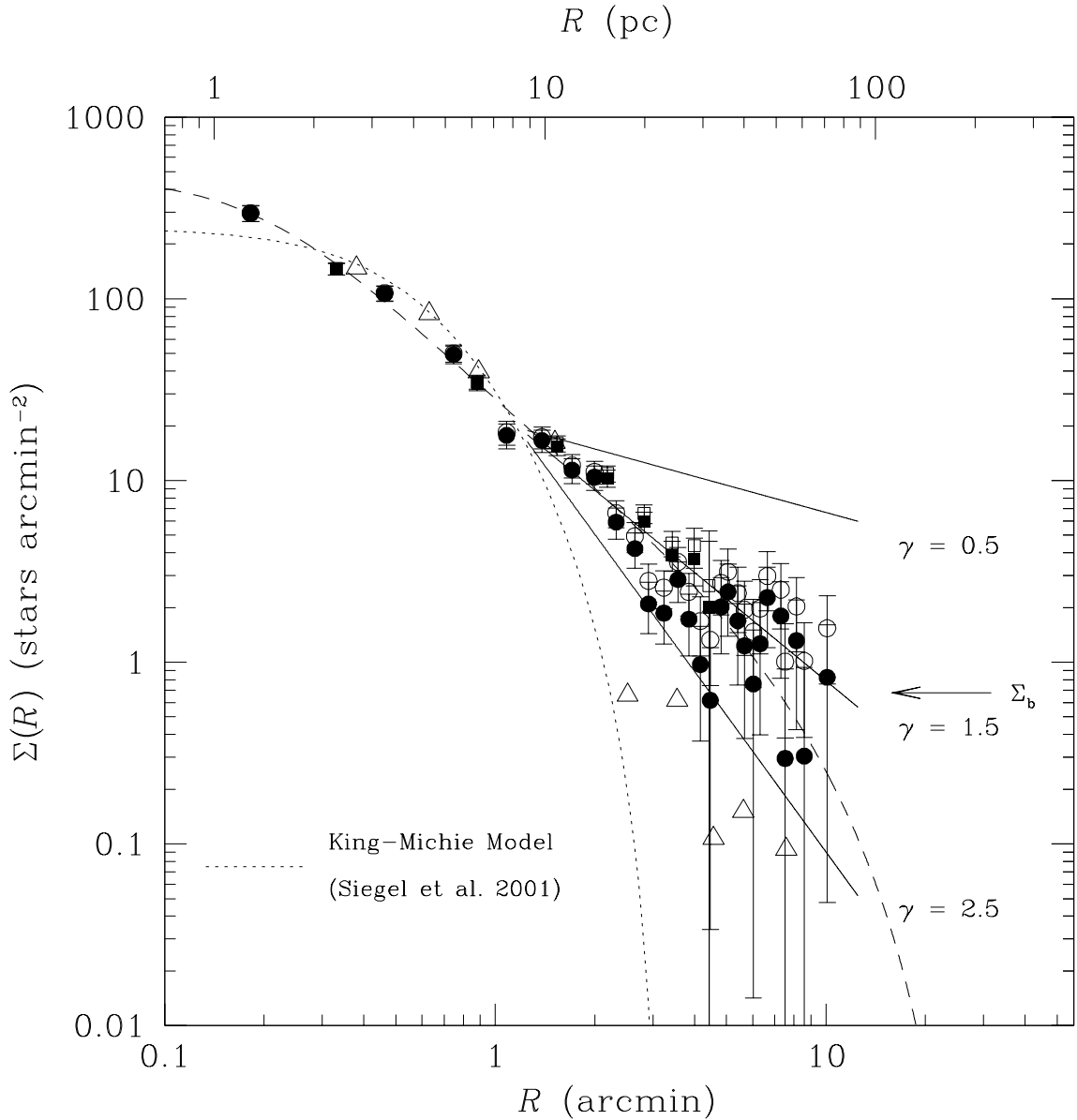}
\caption{Surface density profile for Palomar 13 based on our Keck and
CFHT photometry (squares and circles, respectively). Open symbols show the surface density 
profile before background subtraction; filled symbols show the profile found after 
subtracting backgrounds of ${\Sigma}_b = 0.65$ and 0.62 stars per arcmin$^{-2}$,
respectively, for the Keck and CFHT observations. For comparison, the surface density 
profile of Siegel et~al. (2001) is illustrated by the open triangles, after scaling their 
counts upward by a factor of (629/119) $\simeq$ 5.28 (i.e., the ratio of the number of 
stars used to derive the respective profiles). The dotted curve shows a King model with
concentration parameter $c = 0.7$ and core radius $r_c = 39^{{\prime}{\prime}}$
suggested by Siegel et~al. (2001). The dashed curve shows the formal best-fit 
King-Michie model.  The three dashed lines beginning at $R \simeq$ 1$^{\prime}$ indicate 
profiles of the form predicted by Johnston et~al (2001) for extra-tidal stars: 
$\Sigma \propto R^{-{\gamma}}$ with $\gamma = 0.5,~1.5~{\rm and}~2.5$.
\label{fig5}}
\end{figure}

\clearpage

\begin{figure}
\plotone{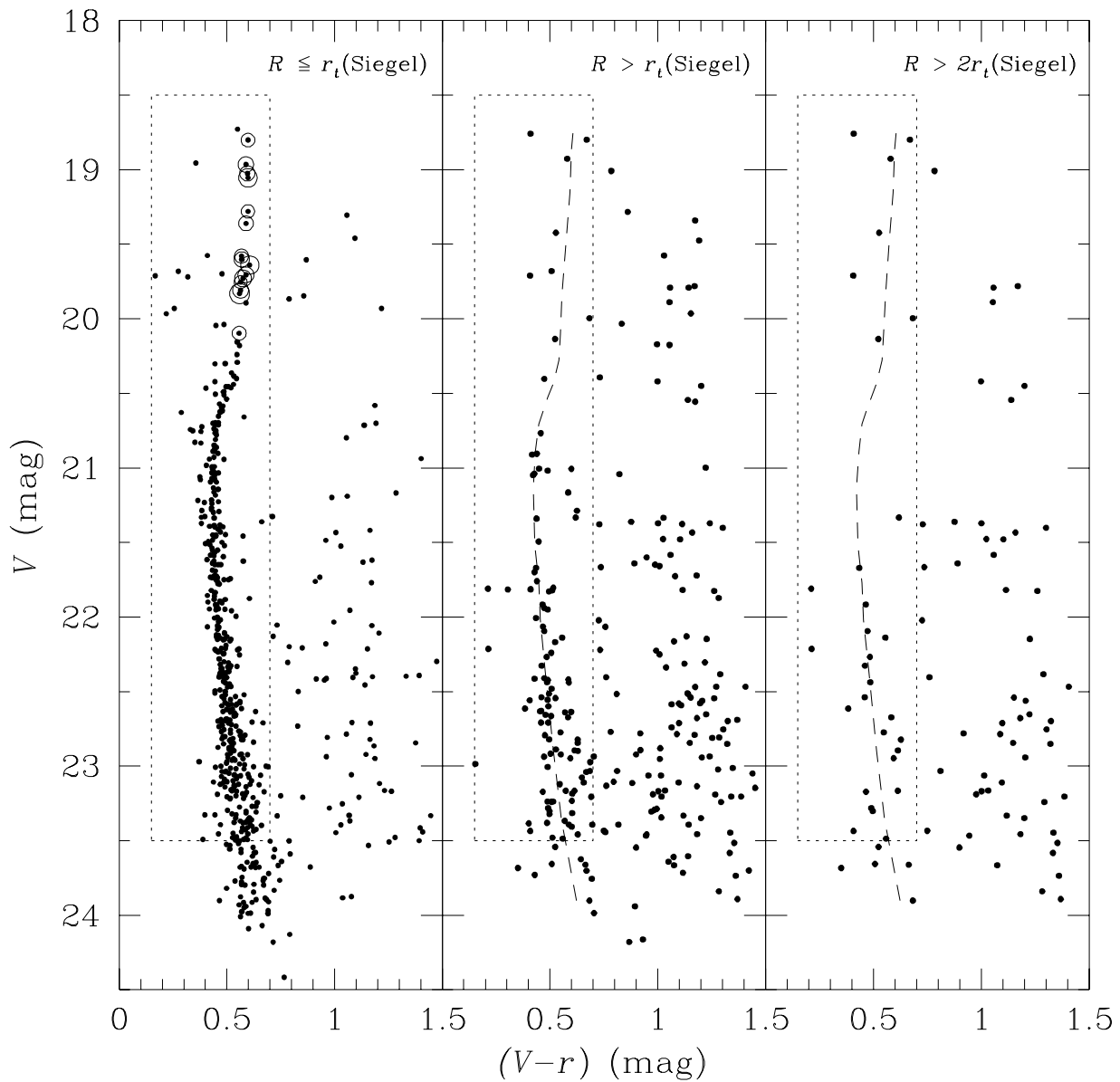}
\caption{Instrumental $V$, $(V-r)$ color magnitude diagrams based on our CFHT images
in three radial regimes. The left panel shows all objects within the canonical tidal
radius, $r_t = 188^{\prime\prime}$, reported by Siegel et al. (2001). The dotted
rectangles shows the magnitude and color limits used to reject probable non-members.
Circled points
indicate probable radial velocity members for which we have reliable CFHT photometry.
All objects located beyond this radius are plotted in the middle panel, where the
dashed curve shows the Palomar 13 fiducial sequence from the previous panel. Objects
at radii greater than {\it twice} the canonical tidal radius are plotted in the right panel.
\label{fig6}}
\end{figure}

\clearpage

\begin{figure}
\plotone{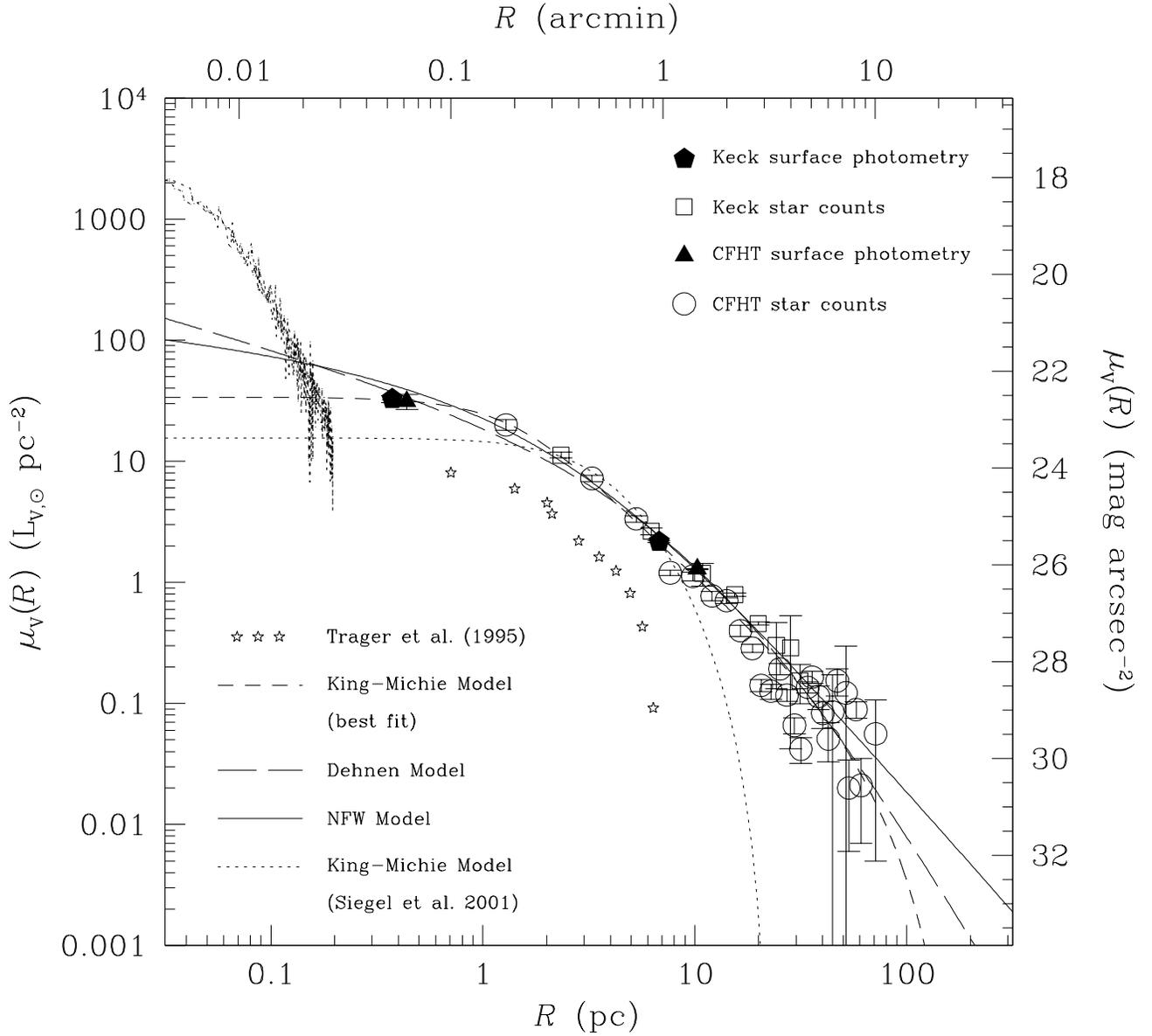}
\caption{$V$-band surface brightness profile for Palomar 13. Pentagons and squares represent
surface photometry and star counts based on our Keck imaging, respectively.
Surface photometry and star counts using our CFHT imaging are indicated by the triangles and 
circles. The King-Michie model which best fits this surface brightness profile is shown by 
the dashed curve. By contrast, the King-Michie model with structural parameters found
by Siegel et al. (2001) from their surface {\it density} profile (see Figure 5) is indicated 
by dotted curve, shifted in the vertical direction. The solid curve shows
the NFW model which best fits the surface density profile, while the long-dashed
curve shows the best-fit Dehnen model.
The dotted curves at the cluster center shows typical {\it stellar} profiles 
on our Keck and CFHT images, scaled to the luminosity of V2 --- the bright
variable which defines the cluster center.
\label{fig7}}
\end{figure}

\clearpage

\begin{figure}
\plotone{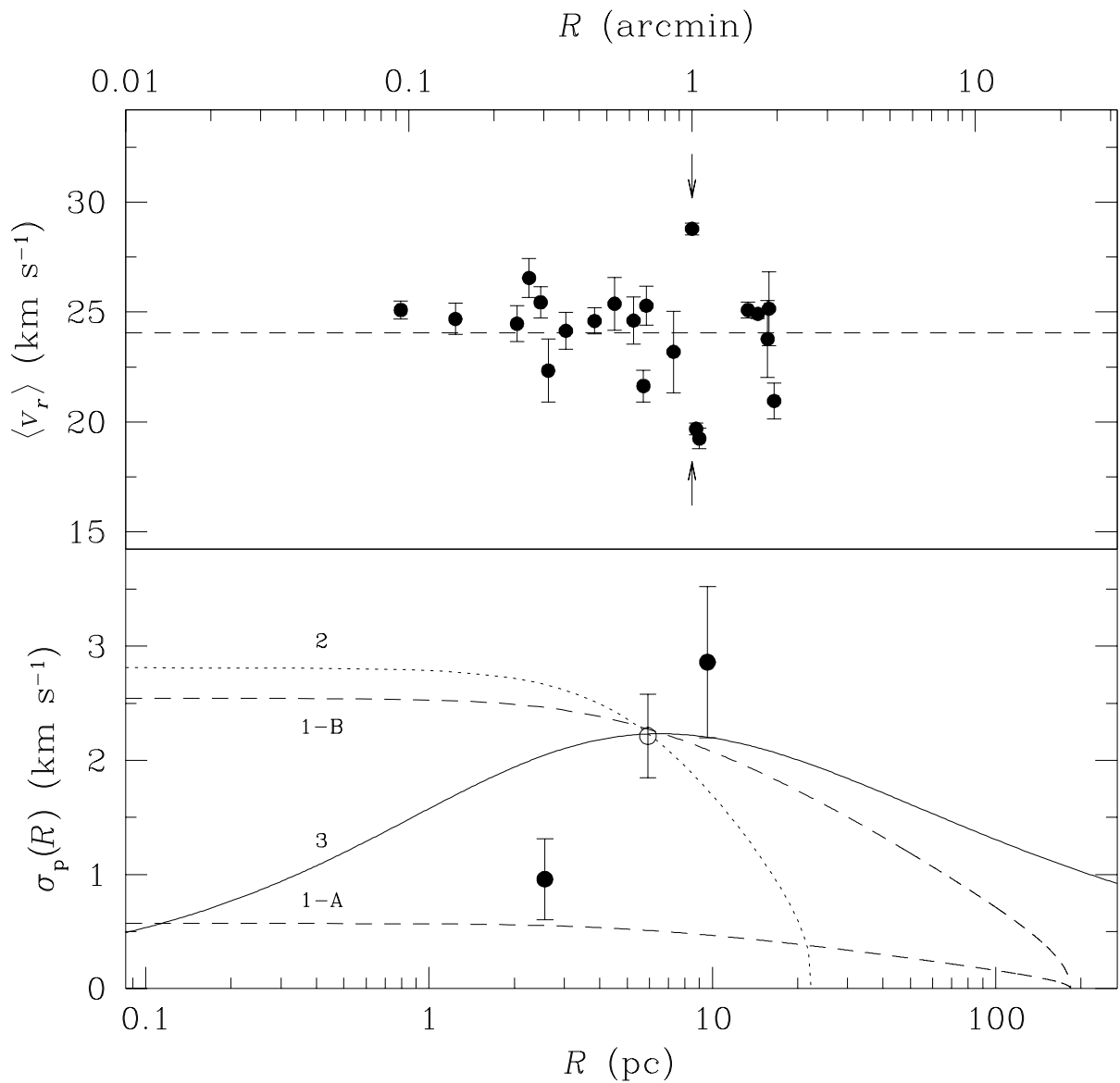}
\caption{(Upper Panel) Heliocentric radial velocity for probable cluster members
plotted against distance from the cluster center (filled circles). The dashed line
indicated the cluster's systemic velocity. The arrows denote the approximate
``break" radius, $R_b \simeq$ 1$^{\prime}$, where the possible excess in the
cluster's surface density profile begins.
(Lower Panel) Velocity dispersion profile for Palomar 13. The open circle shows
the intrinsic velocity dispersion for the entire sample of stars, plotted at the
mean radius of the sample. The two filled circles show the intrinsic
velocity dispersion which is found if the sample is divided into two radial bins
containing 11 and 10 stars each.  The different curves (labeled 1-4) indicate various
model fits to the data; a detailed description of each model is presented in the text.
\label{fig8}}
\end{figure}

\clearpage

\begin{figure}
\plotone{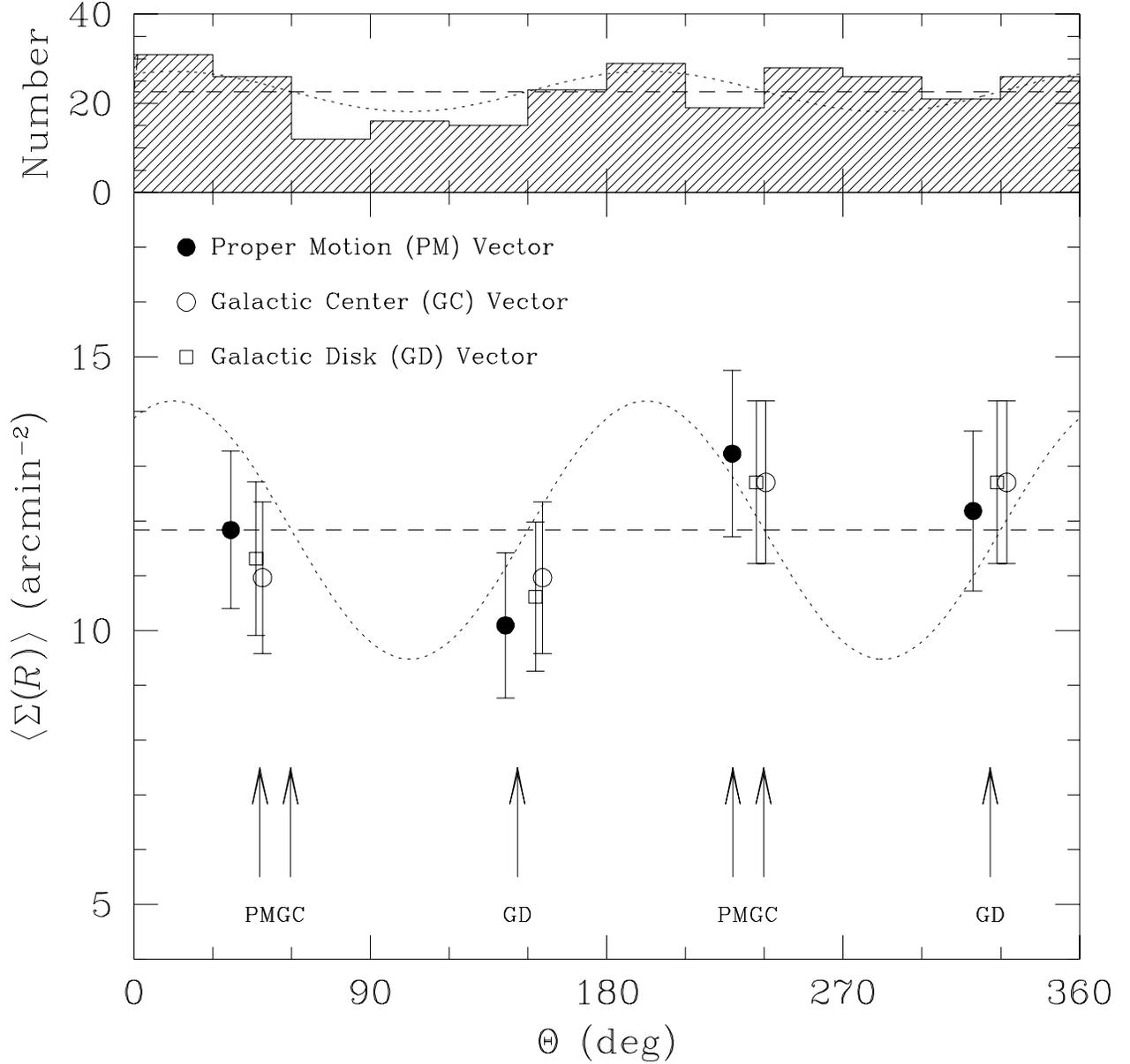}
\caption{(Upper Panel) Histogram of position angles for stars in our photometric database
which are located in the radial range $1^{\prime} \le R \le$ 2\farcm88.
The dashed line shows the mean number of stars per azimuthal bin, while the dotted curve
shows the best-fit sinusoid: $N \propto (4.5\pm1.7)\sin{[2{\Theta} + (60^{\circ}\pm11^{\circ})]}$.
This curve has density maxima at ${\Theta}_p = 15^{\circ}$ and 
$195^{\circ}$. (Lower Panel) Mean stellar surface density in the range 
$1^{\prime} \le R \le$ 2\farcm88 for each of four 90$^{\circ}$ sectors 
centered on cluster center. The dashed and dotted curves correspond to the mean surface 
density and sinusoid described in the upper panel (i.e., with a best-fit amplitude of
2.4$\pm0.9$ arcmin$^{-2}$).
Surface densities are shown 
for three different orientations of the quadrants: in the first case (filled circles), the 
sector boundaries are offset by $\pm$45$^{\circ}$ from the position angle of the vector 
which points in the direction of Palomar 13's absolute proper motion (i.e., the short 
arrow in Figure~\ref{fig1}). In the second case 
(open circles), the boundaries are offset by $\pm$45$^{\circ}$ from the position angle 
of the vector which points in the direction of the Galactic center (i.e., the long arrow 
in Figure~\ref{fig1}). In the final case (open squares), the boundaries are offset by 
$\pm$45$^{\circ}$ from the position angle of the vector which points in the direction 
perpendicular to the Galactic disk (i.e., the intermediate arrow in Figure~\ref{fig1}).
The orientations of these three vectors are indicated by the respective arrows. There
is only marginal evidence for a preferred orientation of Palomar 13 stars in this radial
range.
\label{fig9}}
\end{figure}

\clearpage

\begin{figure}
\plotone{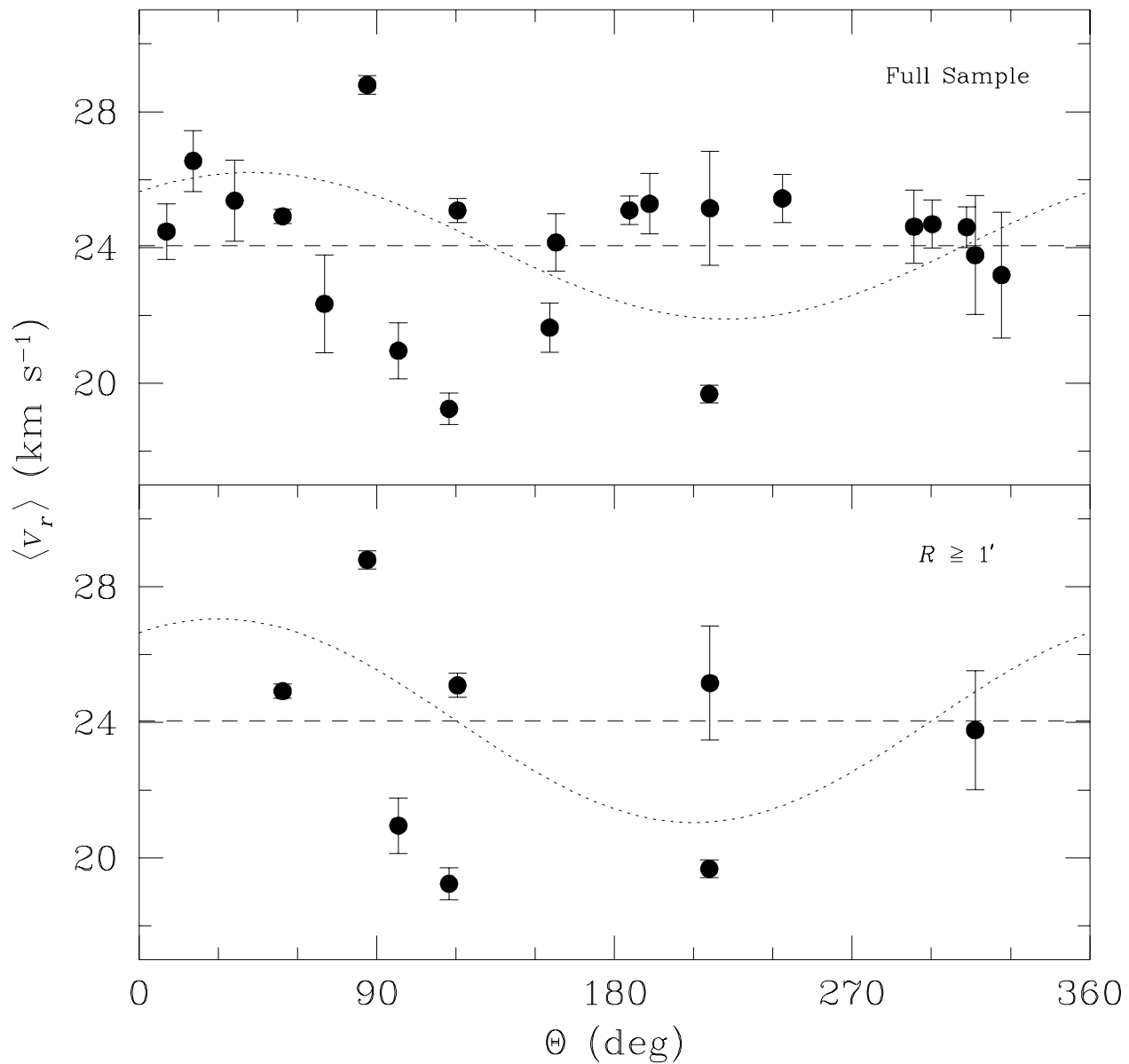}
\caption{(Upper Panel) Mean radial velocity vs. position angle, $\Theta$, for the sample of 21 
probable member stars. The dashed line indicates the systemic velocity, while the dotted
curve shows the best-fit sinusoid of the form 
$\langle v_r\rangle = \langle v_r\rangle_s + {A}sin{(\Theta + \Theta_p)}$, where
$\langle v_r\rangle_s \equiv 24.1$ km~s$^{-1}$.
A weighted least-squares fit yields $A = 2.1\pm1.0$ km s$^{-1}$ and $\Theta_p = 48\pm47^{\circ}$,
while an unweighted fit gives $A = 0.8\pm1.0$ km s$^{-1}$ and $\Theta_p = 104\pm130^{\circ}$.
We conclude that there is no evidence for rotation among the full sample of
Palomar 13 members. (Lower Panel) Same as above, except for the eight stars
located at, or beyond, the ``break radius" of $R_b \sim 1^{\prime}$. The weighted best-fit 
sinusoid is shown by the dotted curve, which has parameters $A = 3.0\pm3.0$ km s$^{-1}$
and $\Theta_p = 60\pm65^{\circ}$. An unweighted fit gives $A = 2.0\pm1.7$ km s$^{-1}$ 
and $\Theta_p = 84\pm44^{\circ}$. Thus, as with the full sample, we find
no evidence for statistically significant rotation for those stars at large radii.
\label{fig10}}
\end{figure}

\clearpage

\begin{figure}
\plotone{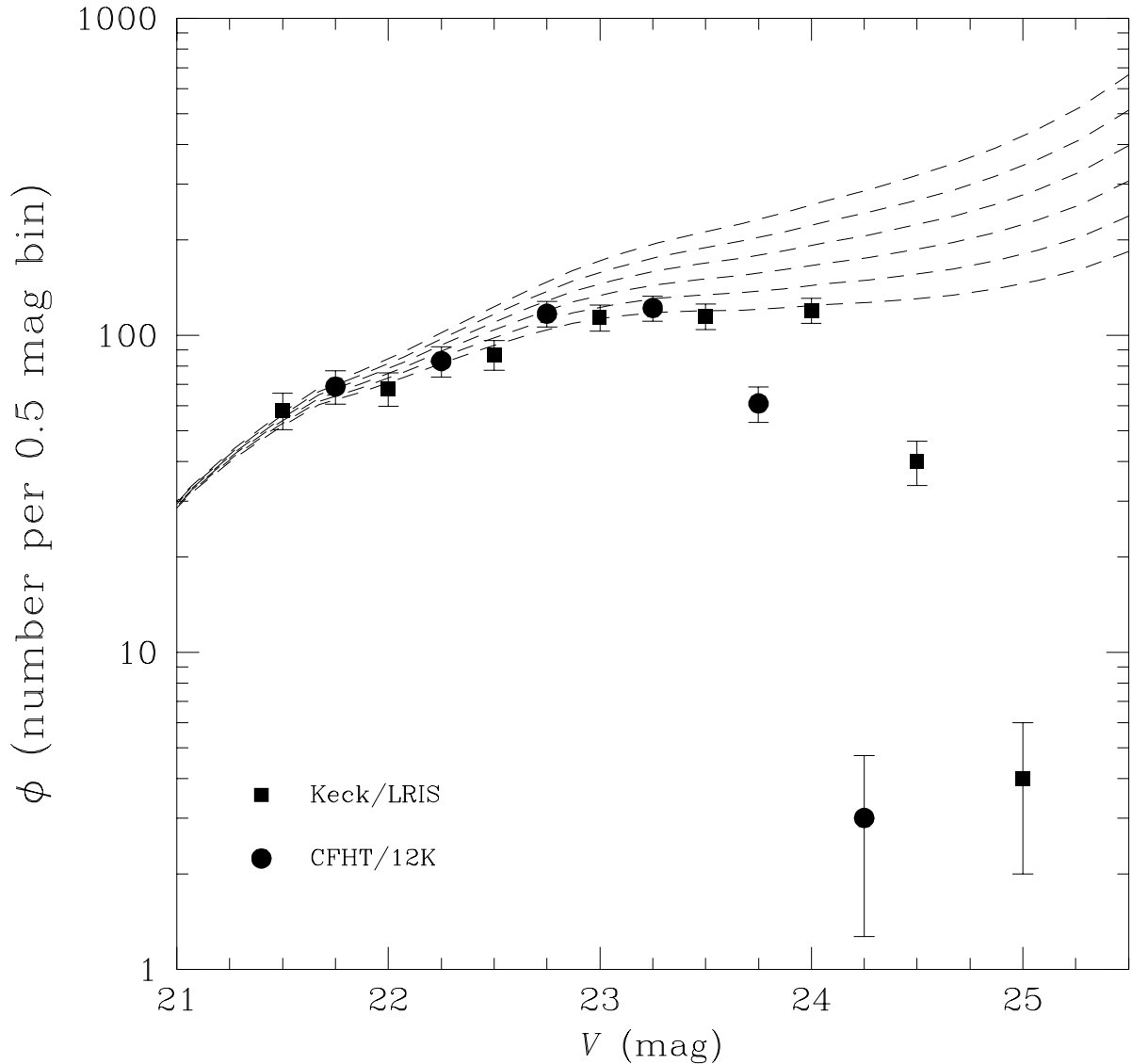}
\caption{Comparison of the observed luminosity function for Palomar 13 with theoretical
luminosity functions for $V \gae 21.5$ mag. Data points from Keck are shown by
the squares; circle indicate measurements from CFHT.
The theoretical luminosity functions have been 
generated from the [Fe/H] = $-$1.78 dex, 14 Gyr isochrone of Bergbusch
\& VandenBerg (1992). Six different values of the mass function exponent, $x$, are shown; from
top to bottom, the dashed curves have
$x$ = 2.5, 2.0, 1.5, 1.0, 0.5 and 0.0. The steep dropoffs at $V \gae 23.5$ mag (CFHT) and 
$V \gae 24$ mag (Keck) are due to photometric incompleteness. Thus, to the completeness limits of our 
data, we find no evidence for a steeply rising luminosity function in Palomar 13.
On the contrary, the observed luminosity function appears quite flat, with $x \sim 0$,
indicating an possible {\it underabundance} of low-mass stars.
\label{fig11}}
\end{figure}

\clearpage

\begin{figure}
\plotone{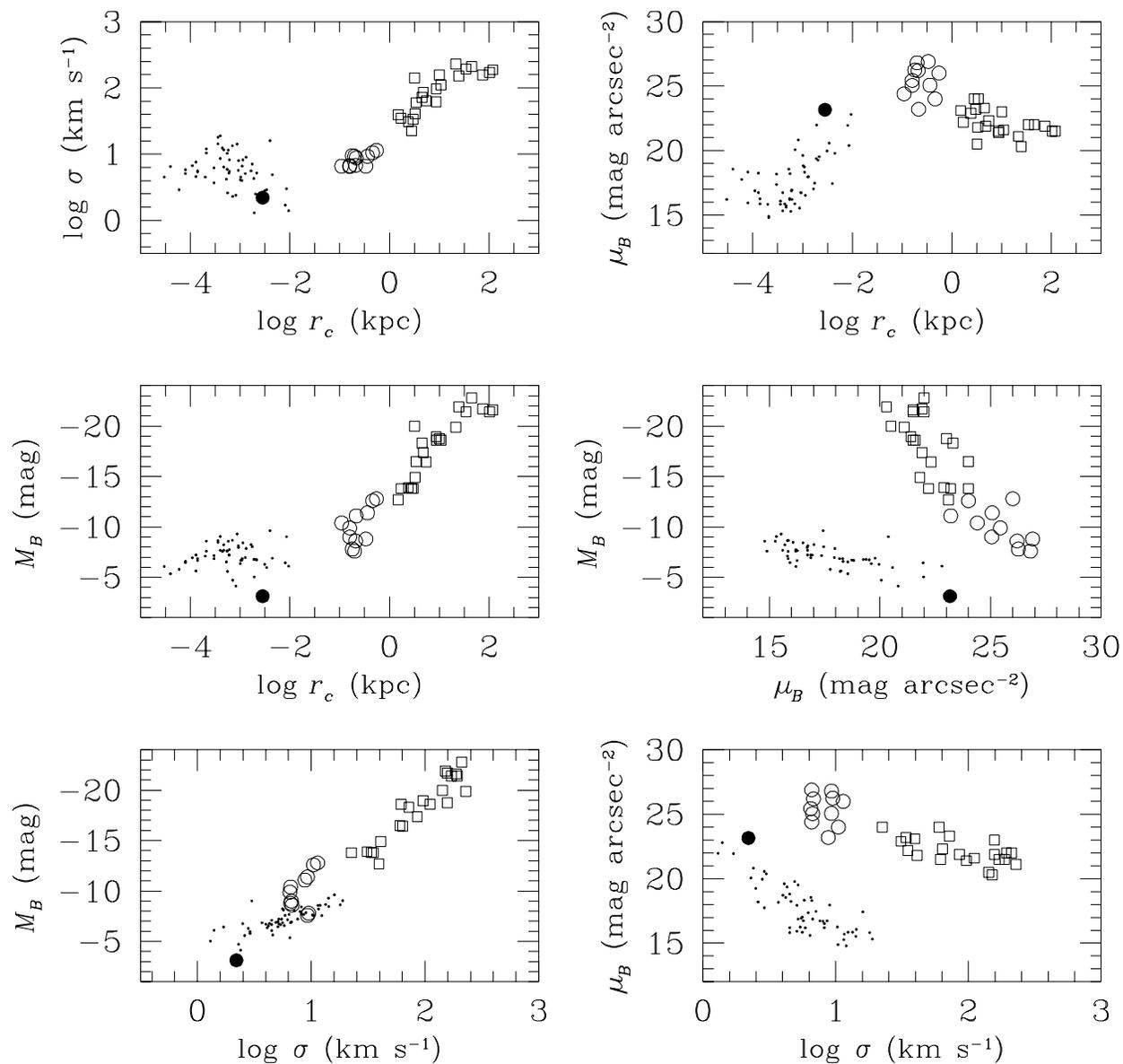}
\caption{Scaling relations for stellar systems. Local Group dwarf spheroidal galaxies 
(data from Mateo 1998) are indicated by open circles. Open squares show the results for 
a sample of dwarf irregulars and spirals taken from C\^ot\'e et~al.  (2000). Galactic 
globular clusters having measured velocity dispersions are shown by the dots 
(Pryor \& Meylan 1993; Harris 1996). Palomar 13 is indicated by the large filled circle.
\label{fig12}}
\end{figure}

\clearpage

\begin{figure}
\plotone{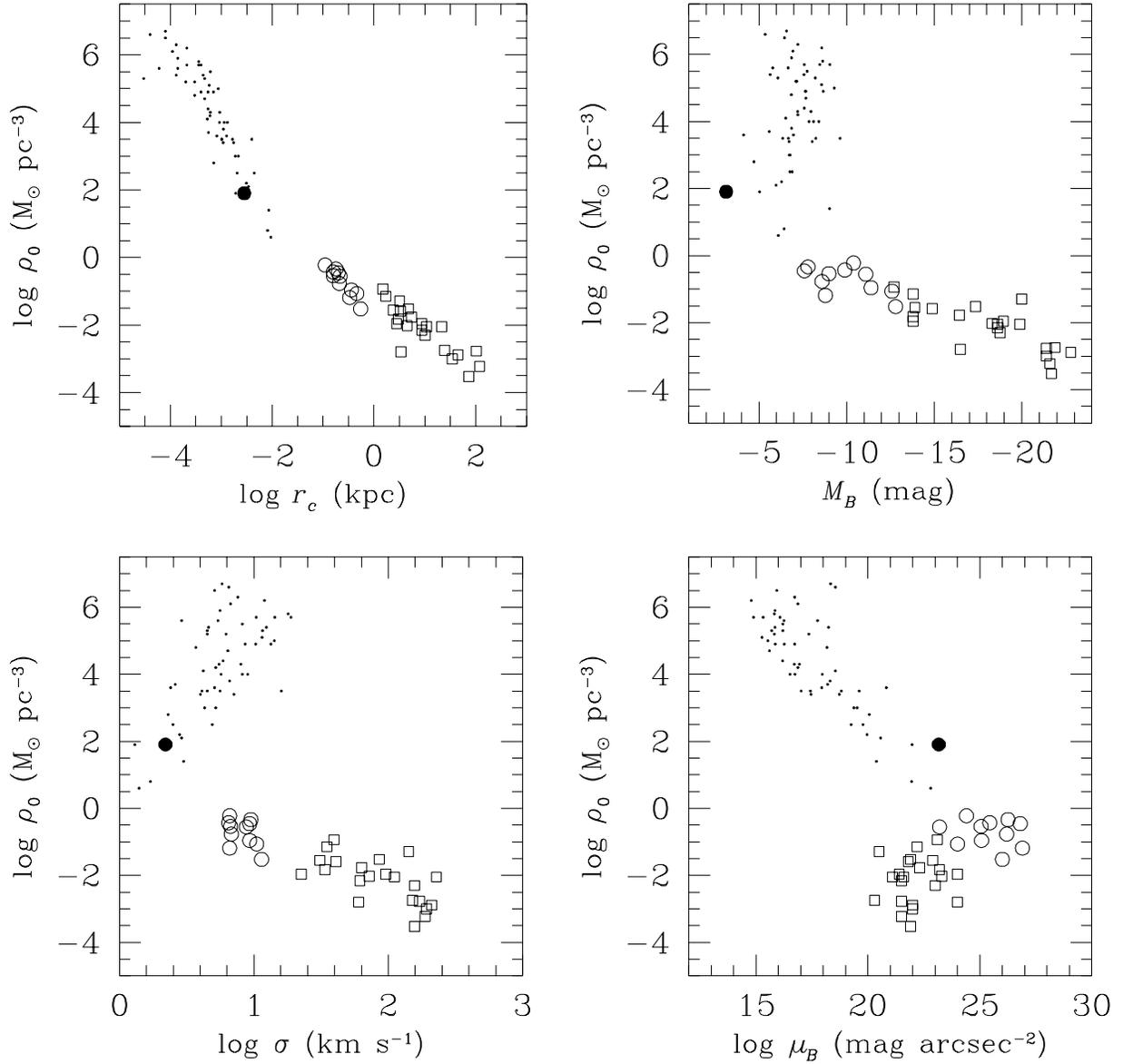}
\caption{Additional scaling relations for galaxies and Galactic globular clusters.
The symbols are the same as in the previous figure.
\label{fig13}}
\end{figure}

\clearpage

\begin{figure}
\plotone{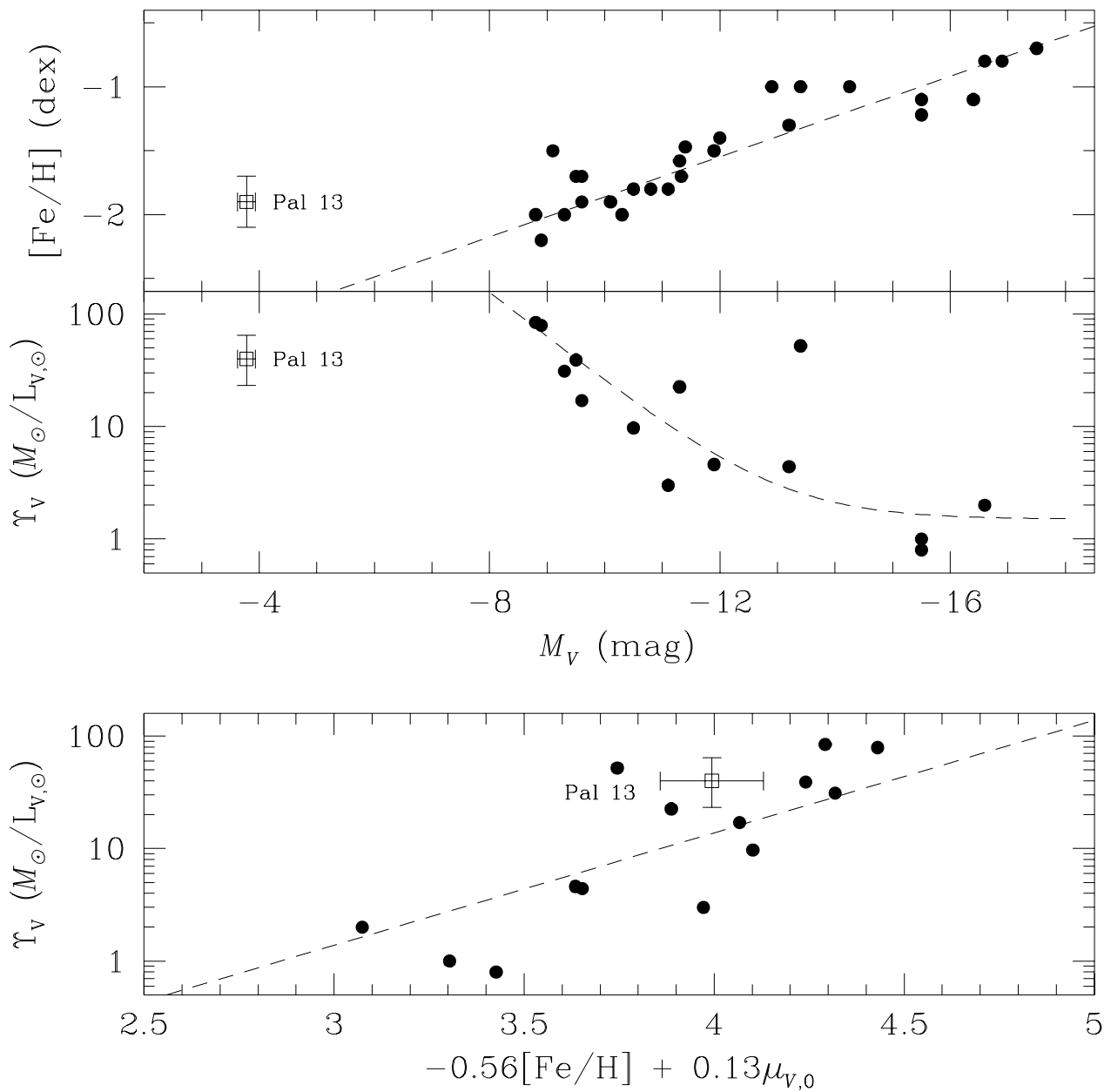}
\caption{(Upper Panel) Metallicity-luminosity relation for dE/dSph galaxies (filled
circles).  Palomar 13 is indicated by the open square. The dashed line shows the best-fit
relation from C\^ot\'e et al. (2000), extrapolated to faint magnitudes. 
(Middle Panel) $V$-band mass-to-light ratio, $\Upsilon_V$, versus absolute visual 
magnitude for Local Group dE/dSph galaxies (see C\^ot\'e et al. 1999). The dashed curve 
shows the relation for dwarf consisting of luminous stellar components with ${\Upsilon}_V = 2$
that are embedded in dark matter halos of mass $M = 2\times10^7~M_{\odot}$. Palomar 13 obeys
neither this, nor the above, empirical scaling relation for dwarf galaxies.
(Lower Panel) Correlation between mass-to-light ratio, metallicity and central $V$-band
surface brightness for Local Group dE/dSph galaxies (filled circles). The dashed line
shows the linear relation proposed by Prada \& Burkert (2001). Palomar 13 is in reasonable
agreement with the proposed correlation.
\label{fig14}}
\end{figure}

\end{document}